\documentclass[aps,floatfix,superscriptaddress,notitlepage,nofootinbib]{revtex4-1}
\usepackage[utf8]{inputenc}
\usepackage{tikz}
\usepackage{amsmath,amssymb,amsfonts,graphics,graphicx,dcolumn,bm,enumerate}
\usepackage{comment,natbib,appendix}
\usepackage{multirow,color}
\usepackage{chngpage}
\usepackage{afterpage}
\usepackage{xcolor}
\usepackage{amsthm}
\usepackage{natbib}
\usepackage{hyperref}
\usepackage[margin=0.8in]{geometry}
\usepackage{epstopdf}
\usepackage{float}
\usepackage{soul}

\newcommand{\isi}
{\affiliation{Economic Research Unit, Indian Statistical Institute, Kolkata 700108, India.}}

\newcommand{\SRM}
{\affiliation{Department of Physics, SRM University-AP, Andhra Pradesh - 522240, India.}}

\newcommand{\cmp}
{\affiliation{Cond.\,Matter Physics, Saha Institute of Nuclear Physics, Kolkata 700064, India.}}

\newcommand{\raghunathpur}
{\affiliation{Department of Physics, Raghunathpur College, Raghunathpur, Purulia - 723133, India.}}

\begin{document}

\title{Sandpile Universality in Social Inequality: Gini and Kolkata Measures}

\author{Suchismita Banerjee}
\email[Email: ]{suchib.1993@gmail.com}
\thanks{corresponding author}
\isi

\author{Soumyajyoti  Biswas}
\email[Email: ]{soumyajyoti.b@srmap.edu.in}
\SRM

\author{Bikas K. Chakrabarti }%
 \email[Email: ]{bikask.chakrabarti@saha.ac.in}
 \cmp \isi 
 
\author{Asim Ghosh}
\email[Email: ]{asimghosh066@gmail.com}
\raghunathpur
  
\author{Manipushpak Mitra}
\email[Email: ]{mmitra@isical.ac.in}
\isi

\begin{abstract}
Social inequalities are ubiquitous and  evolve towards a universal limit. Herein, we extensively review the values of inequality measures, namely the Gini ($g$) index and the Kolkata ($k$) index, two standard measures of inequality used in the analysis of various social sectors through data analysis. The Kolkata index, denoted as $k$, indicates the proportion of the `wealth' owned by $(1-k)$ fraction of the `people'. Our findings suggest that both the Gini index and the Kolkata index tend to converge to similar values (around $g=k \approx 0.87$, starting from the point of perfect equality, where $g=0$ and $k=0.5$) as competition increases in different social institutions, such as markets, movies, elections, universities, prize winning, battle fields, sports (Olympics), etc., under conditions of unrestricted competition (no social welfare or support mechanism). 
In this review, we present the concept of a generalized form of Pareto's 80/20 law ($k=0.80$), where the coincidence of inequality indices is observed. The observation of this coincidence is consistent with the precursor values of the $g$ and $k$ indices for the self-organized critical (SOC) state in self-tuned physical systems such as sand piles. These results provide quantitative support for the view that interacting socioeconomic systems can be understood within the framework of SOC, which has been hypothesized for many years. These findings suggest that the SOC model can be extended to capture the dynamics of complex socioeconomic systems and help us better understand their behavior.
\end{abstract}

\maketitle

\section{Introduction}
The distribution of wealth has been a topic of discussion and concern throughout human history. There is nothing more unequal than distribution of wealth. Indeed, no physical quantity can be perceived without sophisticated measuring instruments within the span wealth can vary (roughly nine orders of magnitude). This perception of wealth inequality has led to social discord from ancient times, to the extent that the history of popular struggles is broadly that against wealth inequality. Plato in his dialogue (The Law) describing the ideal settlement Magnesia, recommended the ratio of wealth between the poor and wealthy not to exceed 1:4. However, societies have been far more unequal than that and we would like to review here the universal and extreme level of inequality in our society. 

 For the purpose of our discussion here, we broaden our scope to include study of inequalities in assets earned through competitive means, which can also include revenues earned through movies, citations of publications (of individuals, universities, journals etc.), gains from stock market fluctuations and so on. As we shall see, in all such cases inequalities are   ubiquitous. The consequences are, however, not similar.
Wealth inequalities have a variety of severe social consequences: limiting access to basic needs such as food, housing, healthcare; limiting access to education and thereby limiting opportunity for upward mobility and so on. On the other hand,  inequality in, for example,  movie revenues has very little consequence for the larger society. Therefore, while policies exist to moderate wealth inequalities, in the other cases the dynamics are completely unrestricted. Hence such dynamics give a window into the possible situations as to what could happen to inequalities in the distribution of some form of asset when such assets are sought for by competing entities without any restriction.  
 
We argue, however, that irrespective of the particular context, dynamics of inequality follow broadly universal characteristics. 
It is, therefore, possible to shed light on highly complex situation of wealth inequality by characterizing its universal dynamics.  
These characterization, of course, need quantification of inequality through unambiguous, and possibly intuitive measures. We attempt to do just that in this article. 
We study a novel inequality index, compare it with the other existing inequality indices, characterize the universal dynamics followed by those indices by analyzing real data and the same in some self-organised critical sandpile models, driven by the production of entropy.  

\section{Social Inequality and its measures}\label{sec:2}
Over a century ago (in 1896), Vilfredo Pareto \citep[][]{Pareto} noticed that only 20\% of Italy's people possessed 80\% of the country's wealth. 
He proceeded to conduct surveys in other European nations, where he discovered, to his astonishment, that the same distribution holds.
The Pareto Principle, also called Pareto's 80/20 law, asserts that 20\% of the causes are responsible for 80\% of the outcomes. 
In other words, the principle suggests that a small fraction of the factors contribute to a large majority of the results.
% So, Pareto's 80/20 law, also known as the Pareto Principle, is a principle that states that 80\% of outcomes come from 20\% of causes. 
The Pareto Principle has been used to analyze many different areas, from economics to quality management, and even in personal development. 
In business, it is often used to identify the most important areas for improvement. 
For example, if a company wants to improve customer satisfaction, it can use the Pareto Principle to identify the 20\% of customers who are responsible for 80\% of the complaints.

\begin{figure}[H]
\centering
\includegraphics[width=10cm]{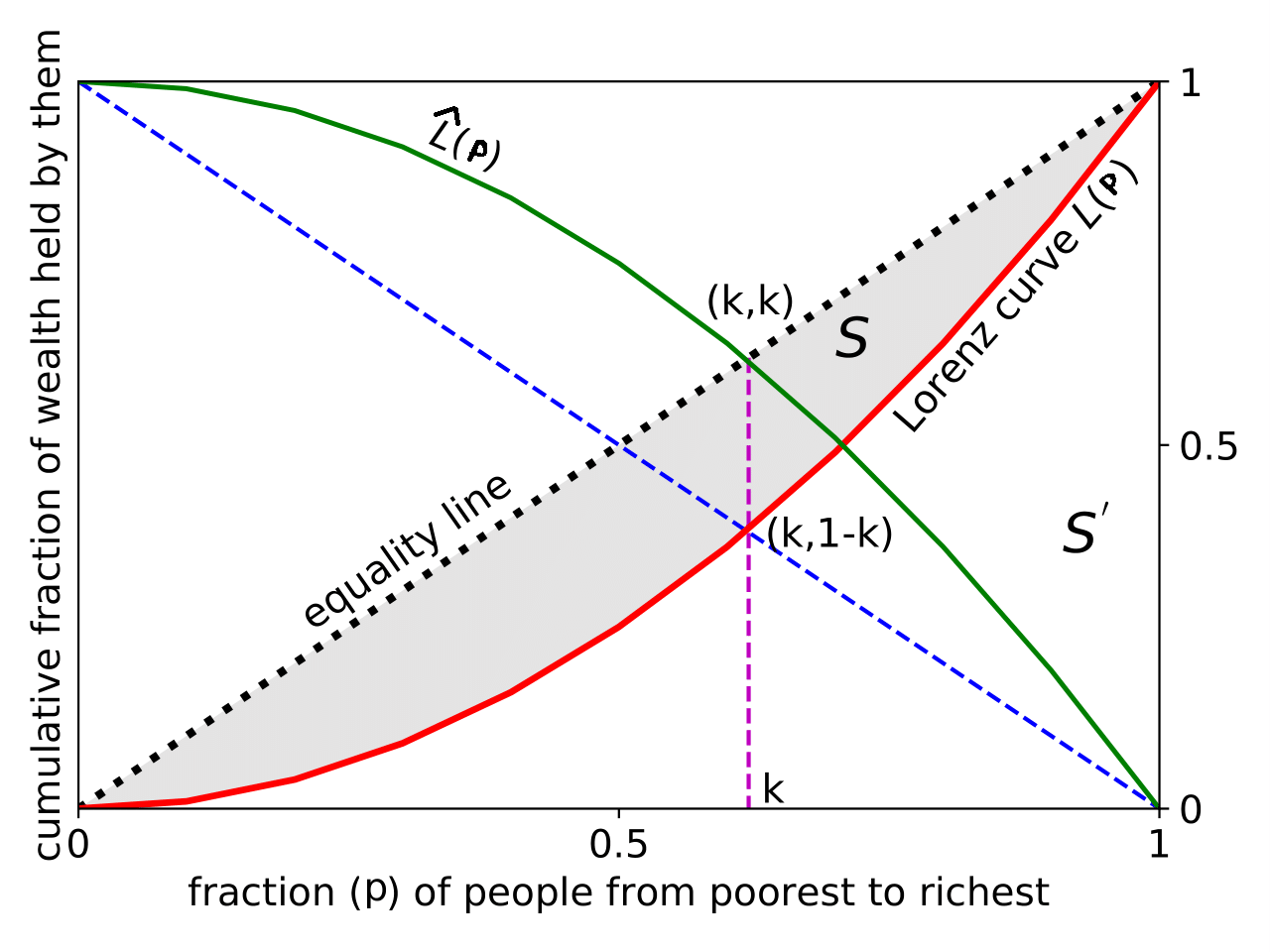}
 \caption{The Lorenz curve or function $L(p)$ (in red) shows the proportion of total wealth owned by a fraction $p$ of people in ascending order of wealth. The black dotted line represents perfect equality where everyone has the same wealth. The Gini index ($g$) is calculated as the area ($S$) between the Lorenz curve and the equality line (shaded region), normalized by the total area under the equality line ($S+S^{'}=\frac{1}{2}$). The complementary Lorenz function $\hat{L}(p) \equiv 1-L(p)$ is shown in green. The Kolkata index ($k$) is determined by the point where the Lorenz curve intersects the diagonal line perpendicular to the equality line. The value of $\hat{L}(k)=1-L(k)$ is equal to $k$, which indicates $k$ is a fixed point of $\hat{L}(p)$ and also gives the proportion of wealth owned by the top $(1-k)$ fraction of the population.}
 % Lorenz curve or function $L(p)$ (in red) represents the fraction of cumulative wealth of the fraction $p$ of people possessing that, when they are arranged from the poorest to the richest. The diagonal (black dotted line) from the origin represents the equality line i.e., what the Lorenz curve would have been if each individual possessed exactly the same wealth. The Gini index ($g$) can be measured by the area ($S$) between the Lorenz curve and the equality line (shaded region), normalized by the total area ($S+S^{'}=\frac{1}{2}$) under the equality line: $g=2S$. The complementary Lorenz function $\hat{L}(p) \equiv 1-L(p)$ is shown by the green line. The Kolkata index ($k$) can be measured by the ordinate value of the intersecting point of the Lorenz curve and the diagonal perpendicular to the equality line. By construction, $\hat{L}(k)=1-L(k)=k$ saying that $k$ is the fixed point of $\hat{L}(p)$ and gives the fraction of wealth possessed by $(1-k)$ fraction of the richest population.}
\label{fig:Lorenz curve}
\end{figure}

Following that, in the year 1905, an American economist by the name of Max Lorenz \citep[][]{Lorenz},\citep[][]{Lorenz1} came up with the Lorenz curve, which is a graphical representation of the distribution of wealth in a society. 
If the population of the society is arranged in the ascending order of their wealth, then
the curve is created by plotting the cumulative fraction of the wealth $L(p)$ possessed by the $p$ fraction of the poorest individuals (red curve shown in Fig.~\ref{fig:Lorenz curve}). 

If wealth were perfectly equally distributed, the Lorenz curve would be a straight line from the origin to the top right corner of the graph (black dotted line in Fig.~\ref{fig:Lorenz curve}). 
In reality, the curve is usually downward-slanting, indicating that a relatively small portion of the population holds a disproportionate share of the wealth.
It is based on the fraction of people who possess wealth less than or equal to a certain amount, denoted as $F(p)$, and the fraction of total wealth possessed by those people, denoted as $L(p)$. 
To illustrate this, consider a society of $N$ individuals with wealth distribution defined by a function $f(y)$, where $y$ represents the wealth of each person. 
The fraction of individuals with wealth less than or equal to $p$ is calculated as the integral of $f(y)$ from 0 to $p$, divided by $N$ as the following,
\begin{equation}
    F(p)= \frac{1}{N} \int_{0}^{p}f(y)\,dy
\end{equation}
The fraction of total wealth possessed by those individuals is calculated as the integral of $yf(y)$ from 0 to $p$, divided by the total wealth of the society ($\mu$) as the following,
\begin{equation}
    L(p)= \frac{1}{\mu} \int_{0}^{p}yf(y)\,dy.
\end{equation}
% Lorenz made a parametric plot of the quantity $L(p)$ verses the quantity $F(p)$. 
% Both $L(p)$ and $F(p)$ are monotone non decreasing, and they increase from zero to one as $p$ goes from zero to infinity. 
% This graph, which has come to be called the Lorenz curve, can be plotted in the unit square, as shown schematically in Fig.~\ref{fig:Lorenz curve}.
Lorenz originally produced a graphical representation of $L(p)$ plotted against $F(p)$ using a parametric method. Both $L(p)$ and $F(p)$ are functions that increase monotonically and continuously from zero to one as $p$ ranges from zero to infinity. The resulting plot, now known as the Lorenz curve, can be displayed within a unit square, which is depicted in a schematic form in Figure~\ref{fig:Lorenz curve}.

Mathematically the Lorenz curve can be formulated in a more compact way as the following,
\begin{equation}
    L(p)=\frac{1}{\mu}\int_{0}^{p}F^{-1}(x)\,dx
\end{equation}
where $F^{-1}(x)=\inf\{y|F(y)\geq x\}$ which means it will take the minimum value of $y$ for which $F(y)\geq x$.

Along with the definition of Lorenz curve, for economic inequality another kind of Lorenz curve is utilized, which is known as `Complementary Lorenz curve' ($\hat{L}(p)$) (green curve shown in Fig.~\ref{fig:Lorenz curve}).
The complementary Lorenz curve is a plot that represents the distribution of wealth in a society in terms of the fraction of the total wealth held by the richest fraction of the population. 
In contrast to the standard Lorenz curve, which shows the fraction of the total population holding a given fraction of the total wealth, the complementary Lorenz curve shows the fraction of the total wealth held by the top fraction of the population.
The complementary Lorenz curve is often used in the study of income and wealth inequality, as it provides a different perspective on the distribution of wealth in a society. 
For example, it can be used to compare the wealth held by the top 1\% of the population to the wealth held by the bottom 99\% of the population.
In a society with a perfectly equal distribution of wealth, the standard Lorenz curve and the complementary Lorenz curve would be the same and would be represented by the 45-degree line.
In a society with a high degree of inequality, the complementary Lorenz curve would be steeper than the 45-degree line, indicating that a large fraction of the total wealth is held by a small fraction of the population.

% \begin{figure}[H]
% \begin{center}
% \begin{tikzpicture}[scale=6]
% \draw (1, 1)--(0, 1)--(0, 0) node [pos=0,above]{D} node [pos=1,below]{A}; 
% \draw [fill=yellow](0, 0)..controls(0.75, 0.25)..(1, 1)--(0, 0); 
% \draw[fill=pink] (0, 0)..controls(0.75, 0.25)..(1, 1) -- (1, 0) -- (0, 0); 
% \draw[green,thick] (0, 0)--(1,0)--(1, 1) node [pos=0,right,black]{B}; 
% \draw[orange,thick](0,1)..controls(0.75, 0.75)..(1, 0);
% \draw [blue,thick] (0, 0)--(1, 1) node[pos=0,below left,black]{(0, 0)}node [pos=1, right,black]{(1, 1)} node [pos=1,above,black]{C}; 
% \draw (0,1)--(1, 0) node [pos=0,left] {(0, 1)} node[pos=1,below ] {(1, 0)}; 
% \draw [red,thick](0,0)..controls(0.75, 0.25)..(1, 1); 
% \draw [->](0.935,0.81)--(1.1,0.81)node [pos=1,right] {Lorenz curve, L(p)};
% \draw[->](0.55,0.55)--(1.1,0.55)node[pos=1,right]{Equality line};
% \draw[->](1,0.3)--(1.1,0.3)node[pos=1,right]{Extreme  inequality line}; 
% \draw [->](0.95,0.15)--(1.1,0.15)node [pos=1,right]{Complementary Lorenz curve, $\hat{L}(p)$}; 
% \draw[dashed](0.688,0)--(0.685,0.316)node [pos=0,below ]{Q}; 
% \draw [black,thick] (0.5,0.5)--(0.685,0.316) node[pos=0,left]{O} node[pos=1,right]{P};
% \end{tikzpicture}
% \caption {The Lorenz and the complementary Lorenz curves.}
% \label{fig:Lorenz curve_1}
% \end{center}
% \end{figure}

Therefore, to a certain extent, the Lorenz curve can be used to measure inequality. While the Lorenz curves contain the full information of the wealth distribution and consequently that of wealth inequality, it is not often convenient to deal with the full distribution function, particularly when a comparative analysis or ranking is required. This calls for some form of a summary statistics to be drawn from the Lorenz curves. 
%Since Lorenz curves can cross one another, the preceding criterion cannot definitively rank them unambiguously.
%The limitations of the Lorenz curve become apparent when two curves cross. 
%Again, an income inequality index is therefore a scalar measure of economic disparities within a population. 
%High income disparity concentrates wealth and shrinks the middle class. 
%Many ways, a large, wealthy middle class improves society. 
%A large, wealthy middle class boosts economic growth, health, education, tax revenue, infrastructure, and social cohesion through fellow feeling. 
%A small middle class and more people outside the middle income group may strain ties between the subgroups on each side of the middle class, causing discontent. 
%Thus, identifying income inequality through several indices is crucial.
For this reason, the Italian statistician and sociologist Corrado Gini in 1912, developed the Gini index \citep[][]{Gini}.
The Gini index is a widely used metric for measuring statistical dispersion in the distribution of wealth or income among residents of a nation. It is commonly used to assess levels of inequality in a society.
To compute the Gini index, the area between the Lorenz curve and the line of equality is divided by the total area under the line of equality. This ratio provides a numerical value that represents the degree of inequality present in the distribution of wealth or income.
%which is a measure of statistical dispersion intended to represent the income or wealth distribution of a nation's residents, and is the most commonly used measure of inequality.
%The Gini index is calculated by taking the ratio of the area between the Lorenz curve and the line of equality to the total area under the line of equality.
In Fig.~\ref{fig:Lorenz curve}, Gini index, $$g = \frac{S}{(S+S^{'})}.$$
Equivalently, the Gini index is twice the
area between the Lorenz curve and the line of perfect equality,
\begin{equation}
    g = 2\int_{0}^{1}(p-L(p))\,dp = 1-2\int_{0}^{1}L(p)\,dp.
\end{equation}
The Gini index ranges from 0 to 1, with 0 representing perfect equality and 1 representing perfect inequality. 
The Gini index is used by economists to measure the income inequality of a nation, and is sometimes used to measure the inequality of other variables, such as wealth distribution and health inequality.
The significance of adopting alternative measures of inequality is commonly recognised because no single summary statistic can capture all characteristics of inequality displayed by the Lorenz curve and Gini index.
Particularly for the Gini index, it is known to be a rather non-intuitive measure i.e., quoting its value does not give an immediate picture of the underlying inequality. Indeed, the problem is deeper than intuition. Multiple Lorenz curves can have the same Gini index value. This means, calculating Gini index does not uniquely define the inequality. 

%%%%%%%%%%%%%%%%%%%%%%%%%%%%%%%%%%%%%%%%%%%%%%%
% Before the industrial revolution, most economic activity was subsistence or barter.
% People farmed, made their own things, and traded with neighbours.
One major factor contributing to economic inequality is socialization, or the tendency for the rich to become richer. 
This can occur because those with wealth have more resources to invest and accumulate more wealth, while those without wealth may struggle to get ahead.
As a result, the gap between the rich and the poor can grow over time.
% Economic socialisation changed in the late 1800s.
The industrial revolution transformed economic interactions.
Shops sold things produced more efficiently by factories.
Economic socialisation followed the industrial revolution.
The automobile and telephone made travel and communication easier.
Economic socialisation changed during the Great Depression.
As employment and wages disappeared, people had to find other methods to make money. 
To help people cope with the economic crisis, unemployment insurance and welfare benefits were created.
World War II reshaped economic socialisation by the mid-$20^{th}$ century.
The US standard of living rose as more commodities were available.
Wealth spurred consumerism and the service industry.
% 1950 was economically unequal.
Socialization during this time gave more people the chance to rise economically and partake in the prosperity.
The wealthy had more resources, jobs, and education than the poor.
For these, global Gini ($g$) index value is increasing and in 1980-90, this value remained at $g \simeq 0.65$ (see \citep[][]{Globalization}), i.e, the top 35\% of the population earned almost 65\% of the overall income.   
The US tax policy favoured the wealthiest, allowing them to pay lower rates than the middle- and lower-classes, increasing wealth disparity.
% In 1950, class, race, and gender shaped socialisation.
% The wealthy used their means to get greater possibilities, whereas the poor were generally excluded.
% This increased wealth inequality.
% The Gini index had a value of less than 0.5 at this period of time.
Economic policies began to impact residents' economic behaviour at this time.
Taxation, subsidies, and other economic interventions did this.
% The public sector's share of GDP rose to 25\% in 1950.
% Citizens received unemployment relief, health care, and pensions from governments.
% Consumerism soared in the 1950s.
% Credit, rising consumer goods affordability, and widespread availability drove this.
% The economy grew rapidly as firms produced more goods and services to meet rising consumer demand.
% Economic socialisation in 1950 shaped the modern economy.
% Consumption soared, the public sector expanded, and welfare payments and public services were introduced.
% These improvements gave many people economic security and shaped the current economy.
In conclusion, economic inequality is a complex issue with many factors contributing to its persistence and growth. 
While tools like the Lorenz curve and Gini index help to understand and quantify inequality, addressing the root causes of this issue will require a more comprehensive approach.

In recent years, there has been a growing concern about the increasing concentration of wealth in the hands of a small percentage of the population. 
In 2011, the `Wall Street Occupy' protest \citep[][]{Protest} was a response to this trend, with participants calling for greater economic equality and a fairer distribution of wealth.
The movement consisted of a campaign of civil resistance, with protesters occupying public spaces in cities across the country to demand changes to the economic and political system. 
The movement began on September 17th, 2011, when a group of protesters entered Zuccotti Park in New York City and began an ongoing occupation of the park. 
The protests quickly spread to other cities in the United States and around the world, inspiring similar protests in countries such as the United Kingdom, Spain, and Greece. 
The movement has been largely credited with helping to spark the global Occupy movement, which has seen occupations in many countries. 
The movement also served as a catalyst for a wide range of progressive issues, including income inequality, corporate greed, and the power of the financial sector.

The protesters argued that, unless something is done to address economic inequality, 99\% of the wealth would soon be possessed by just 1\% of the population.
The most famous slogan of the Occupy Wall Street movement was ``We are the 99\%''. 
This slogan was used to highlight the inequality that exists between the wealthiest 1\% of the population and the other 99\%.
The slogan was meant to emphasize that the economic system is broken and that the wealthy 1\% are exploiting the resources of the other 99\%. 
The slogan has since been adopted by various movements around the world, and is still in use today.

Now the question is, really, the disparity is this great? 
In reality, is it possible to accurately determine what amount of population possesses exactly what amount of wealth?
To find the answers of these queries, in 2014, a new social inequality measure is introduced, named as $k$-index or Kolkata index \citep[][]{k-index},\citep[][]{k-index1},\citep[][]{k-index2}, giving the $k$ fraction of wealth, citations or vote shares  possessed, attracted or obtained by the richer or successful $(1-k)$ fraction of people, papers or election candidates respectively (see also \cite{Subramanian} and references therein).
In Fig.~\ref{fig:Lorenz curve}, k point is the $k$-index.
Mathematically, we can say that, the $k$-index for any income distribution is defined by the solution to the equation,
\begin{equation}
    k + L(k) = 1.
\end{equation}
Extensive data analysis implies that often a $k$-value of more than the Pareto value 0.80 across competitive economies (where the welfare measures are being withdrawn), citation shares in top-ranked universities or successful individual scientists or vote shares in vibrant democratic elections has perhaps continued for ages beyond our notice.  
Indeed, observations suggest that the inequality measure ($k$) becomes as high as 0.86, which is more than the Pareto value ($k$ = 0.80) but less than many apprehended limits or conjectures, like that in the `Occupy Wall Street' slogan ($k$ = 0.99). 
In short, data-analysis suggests that almost 14\% of people, papers, election candidates or even wars  possess, attract, capture or cause about 86\% of the wealth, citations, votes or deaths respectively.
For example, study shows that about 12\% of the 2386 publications (books, documents, letters, etc) by Karl Marx, as collected in his Google Scholar page maintained by the British National Library, accounts for about 88\% of the total 424810 citations as of today collected in that page.

Further, in 2016, Watkins et. al. wrote a review paper \citep[][]{Watkins} on evolution of social systems to Self-Organized Critical state.   
Social systems evolve in complex and unpredictable ways.
However, many theorists have suggested that social systems tend to move towards a state of self-organizing criticality, or SOC state (see \citep[][]{Per1},\citep[][]{Per2}). 
This is a state in which the system is balanced between order and chaos, allowing for a dynamic and adaptive response to changing environmental and social conditions. 
Such self-organization can result in emergent properties, such as the emergence of group norms, the development of new practices or technologies, and the emergence of social networks that can facilitate collective action. 
Ultimately, the exact form that a social system will take is dependent upon a variety of factors, including the resources available and the environmental and social conditions.
Examples of SOC include earthquakes, forest fires, and the stock market.

Therefore, Piketty highlights in his book published in 2017 \citep[][]{Pickety} the fact that the wealth of the top 10\% of the population is continuously expanding at an alarming rate.
Piketty argues that there is a tendency for the rate of return on capital to be higher than the rate of economic growth, which leads to the concentration of wealth among a small number of individuals. 
He suggests that this concentration of wealth poses a threat to democratic societies, as it can lead to social and political instability.

After that in 2020, extensive data analysis shows that \citep[][]{Suchi}, as competition increases in various social institutions, including as markets, universities, and elections, the values of the two general inequality indices, the Gini ($g$) and Kolkata ($k$) indices, approach one another. 
It is further demonstrated that in unrestricted competitions, these two indices equalise and stabilise at a certain value ($k_{max}\simeq 0.87 \simeq g_{max}$). 
We suggest seeing this coincidence of inequality indices as a broader application of Pareto's 80/20 rule ($k=0.80$).
Additionally, the synchronicity of the inequality indices noted here is strikingly comparable to those of self-organized critical (SOC) systems previously discussed. 
The findings outlined the following sections provide quantitative evidence for the long-held hypothesis that interacting socio-economic systems can be seen within the context of SOC systems. 

Few social inequality indicators have been explored in relation to the sub critical dynamical properties (measured in terms of the avalanche size distributions) of some SOC models as their respective stationary critical states approach. 
% In 2022, it has been observed \citep[][]{Manna} that these inequality measures (specifically the Gini and Kolkata indices) exhibit nearly universal values for vastly different models like: the Bak–Tang–Wiesenfeld (BTW) sand-pile, the Manna sand-pile, the quenched Edwards–Wilkinson interface (EW model), and the fibre bundle interface (FBM).
% These findings show that SOC systems share a high degree of commonality with regard to these inequality indicators.
% A comparison to similar earlier results in the data of socioeconomic systems with unrestricted competitions suggests the emergence of inequality due to the likely proximity to SOC states i,e. for SOC models $k$-index is also $\simeq 0.86$.
A recent study \citep[][]{Manna} (in 2022) has observed that different models of self-organized criticality (SOC) - including the Bak-Tang-Wiesenfeld (BTW) sand-pile, Manna sand-pile, quenched Edwards-Wilkinson interface (EW model), and the fiber bundle interface (FBM) - all display similar precursor to SOC state values of the inequality measures, such as the Gini and Kolkata indices. These results suggest that SOC systems share a high degree of commonality when it comes to indicators of inequality. Additionally, comparing these findings to similar results from socioeconomic systems with unrestricted competitions, it appears that inequality may emerge due to proximity to SOC states. Specifically, the $k$-index for SOC models appears to be around $\simeq 0.86$. These observations provide further evidence for the universality of inequality measures across various physical and socioeconomic systems.
We have parallelly demonstrated numerically that the cluster or avalanche size distributions in the various SOC models of self-tuned physical systems (also argued to model all the social systems, see e.g. \citep[][]{Watkins}) do reach a similar $k$-value as the respective SOC points are approached, indicating that as one approaches the  SOC point, about 86\% of the avalanche mass are carried away by 14\% of the avalanches.
It is exactly similar to the inequalities that we observed by mathematically as well as empirically for different socio-economic systems. 

\begin{figure} [H]
    \centering
    \includegraphics[width=13cm]{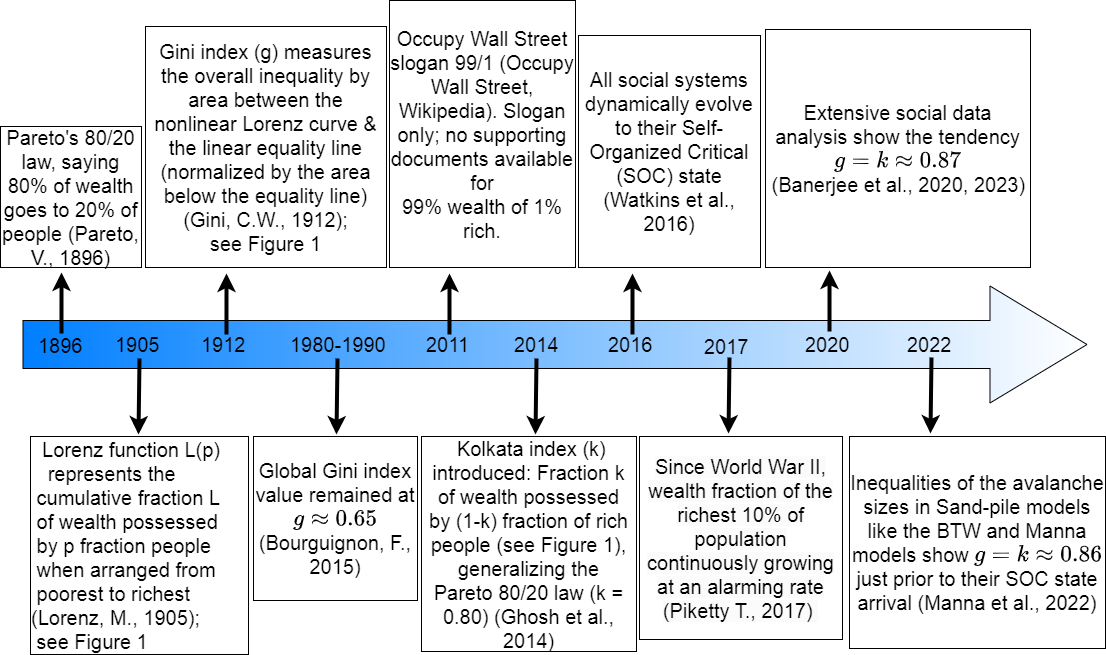} \caption{Timeline of the evolution of social inequality measures since 1896 and their universal convergence to those for sand pile models prior to their respective self-organized critical (SOC) points. We start the timeline from 1896 with the work of Pareto \cite{Pareto} and subsequent developments in 1905 by Lorenz \cite{Lorenz} and Gini \cite{Gini} in 1912. Then we observe the consistency of the Gini index ($g$) for a decade-span 1980-1990 \cite{Globalization}. Subsequent protest happened in 2011 at the Wall Street for the advection of the majority portion of the entire wealth in the hands of very few people \cite{Protest}. In 2014, Kolkata index ($k$) was introduced as another measure of inequality in the wealth distribution \cite{k-index}. In 2016, Watkins and others proposed that all social systems evolve towards the respective SOC state \cite{Watkins}. Piketty (2017) pointed out forcefully about the continuous growth of the wealth of top 10\% of the people \cite{Pickety}. In the year 2020, the work of Banerjee and others reported that the inequality of the social systems has a tendency to evolve at a point of $g=k\approx 0.87$ \cite{k-index2,Suchi}. In 2022, Manna and others showed numerically that many physical SOC systems show $g=k\approx 0.86$ just preceding the SOC points in the respective systems \cite{Manna}.In this review, the figures and tables are arranged with self-contained captions in an attempt to provide readers with an overview of our motivation and the main results presented the introductory
and concluding sections (15~figures and 13 tables and their captions).}
\label{fig:my_label}
\end{figure}

In Fig.~\ref{fig:my_label}, we can see a schematic representation of this timeline story since 1896 at a glance as discussed above.

\section{Calculating the inequality indices}\label{sec:3}
In the last section we have defined what Lorenz function is and also the two inequality measures: Gini and Kolkata indices that summarizes the inequality statistics of a society. In this section, we will proceed to calculate these quantities and try to find analytical relations between them. 

\subsection{\textbf{Properties of the Lorenz curve}}
With the definition of Lorenz curve, as mentioned before, we can enumerate several of the properties that such curve must follow:
\begin{enumerate}
\item The range of Lorenz curve is from $\{F(p),L(p)\}\equiv\{0,0\}$ to $\{1,1\}$.\\ 
\textit{Proof:}
From Eq.~(3), we see that, at $p = 0$, $F(0) = 0$ and $L(0) = 0$. Similarly, for $p = 1$, we have, $F(1) = 1 $ and $L(1) = 1$. Hence, as $p \in [0,1]$, Lorenz curve always range from $\{F(p),L(p)\}\equiv\{0,0\}$ to $\{1,1\}$. (proved). 
\item Lorenz curve is concave and monotonically increasing function of wealth.\\
\textit{Proof:}
\begin{equation}
\frac{d\,L(p)}{d\,p} = \frac{f\,p}{\mu}
\end{equation}
and,
\begin{equation}
\frac{d\,F(p)}{d\,p} = \frac{f}{N}    
\end{equation}
so the slope of the Lorenz curve is given by,
\begin{equation}
\frac{d\,L(p)}{d\,F(p)} = \frac{N}{\mu}\,p   
\end{equation}
As $p \in (0,1)$ and always increasing, from Eq.~(6), one can always say that the slop of the Lorenz curve is always increasing. Hence, Lorenz curve is concave up and monotonically increasing function of wealth. (proved).
\item Lorenz curve for a society where each person possess equal amount of wealth is a diagonal.\\
\textit{Proof:} 
If each person of the society possess equal amount of wealth, wealth distribution follows a Dirac delta function as,
\begin{equation*}
    f(y)=\delta (y-y_{0})
\end{equation*}
Now, 
\begin{equation*}
\begin{split}
    F(p) & = \frac{1}{N}\int_{0}^{p}f(y)\,dy\\
         & = \frac{1}{N}\int_{0}^{p}\delta(y-y_{0})\,dy.
\end{split}
\end{equation*}
So,
\begin{equation*}
    F(p)=
    \begin{cases}
    0 \qquad \text{if}\,p<y_{0} \\
    1 \qquad \text{if}\,p\geq y_{0}
    \end{cases}
\end{equation*}
Again,
\begin{equation*}
    L(p)= 
    \begin{cases}
    0 \qquad \text{if}\,p<y_{0} \\
    1 \qquad \text{if}\,p\geq y_{0}
    \end{cases}
\end{equation*}
Hence, we can see that, $L{p} = F(p)$ i.e, Lorenz curve is the diagonal line for this particular case. (proved).
\item The upper limit of the Lorenz curve is bounded by the equality line.\\
\textit{Proof: }
Now, the second derivative of Eq.~ (6), we have,
\begin{equation*}
   \frac{d^{2}L}{dF^{2}} = \frac{N}{\mu} \left(\frac{d\,F}{d\,y}\right)^{-1} = \frac{N}{\mu}\frac{N}{f}\geq 0
\end{equation*}
Therefore, one can conclude from the above exercise that the Lorentz curve can never reach beyond the diagonal.
Moreover owing to the fact that $L=F=0$ when $p=0$ and $L=F=1$ as $p\to\infty$. In addition, the concave up topology of the Lorentz curve indicate that it is bounded by the diagonal (also known as Egalitarian Line).
\end{enumerate}

\subsection{\textbf{Exemplary calculations of Lorenz curve}}
Here we show calculation of Lorenz curve for some simple wealth distribution functions albeit continuous and discrete. 
\begin{enumerate}
    \item \textit{Uniform wealth distribution: } Let us examine a society in which the distribution of income is uniform over a finite range of values within the interval $[a,b]$, with $0<a<b<1$. The corresponding probability density function is given by $f_{u}(x) = \frac{1}{(b-a)}$, and the cumulative distribution function is $F_{u}(x) = (x-a)/(b-a)$ for all values of $x$ within $[a,b]$. Applying Eq.~(3), we obtain the following Lorenz curve for this distribution:
    \begin{equation}
        L_{u}(p) = \frac{1}{\mu_{u}}\int_{0}^{p}\{a+(b-a)q\}\,dq = p\left[ 1-\frac{(b-a)}{(a+b)}(1-p)\right],
    \end{equation}
   The distribution has a mean of $\mu_{u} = \frac{(a+b)}{2}$, and $F_{u}^{-1}(q) = a+(b-a)q$. It is worth noting that if $a=0$, the Lorenz curve simplifies to $L_{\Bar{u}}(p) = p^{2}$.
    \item \textit{Exponential wealth distribution: }
    Let us consider an exponential income distribution, characterized by the probability density function $f_{E}(x) = \lambda e^{-\lambda x}$, where $\lambda > 0$, and the cumulative distribution function $F_{E}(x) = 1-e^{-\lambda x}$ for all $x \geq 0$. The mean of this distribution is given by $\mu_{E} = \frac{1}{\lambda}$, and $F_{E}^{-1}(q) = -(\frac{1}{\lambda})\ln{(1-q)}$. The Lorenz curve for this distribution is therefore given by:
\begin{equation}
    L_{E}(p) = \int_{0}^{p}-\ln{(1-q)}\,dq = p - (1-p) \ln{\left(\frac{1}{1-p}\right)}.
\end{equation}
    \item \textit{Pareto wealth distribution: } Let us now consider a society with a Pareto-like income distribution. The probability density function for this distribution is given by $f_{P,\alpha}(x) = \frac{\alpha (m)^{\alpha}}{(x)^{\alpha+1}}$, and the cumulative distribution function is $F_{P,\alpha}(x) = 1-(\frac{m}{x})^{\alpha}$, where $m > 0$ is the minimum income, $\alpha > 0$, and the probability density and cumulative distribution functions are defined for all $x \geq m$. The mean of this distribution is $\mu_{P} = \frac{\alpha m}{(\alpha - 1)}$, and $F_{P,\alpha}^{-1}(q) = m(1-q)^{-(\frac{1}{\alpha})}$, which gives the Lorenz curve as follows:
    \begin{equation}
        L_{P,\alpha}(p) = \frac{(\alpha -1)}{\alpha}\int_{0}^{p}(1-q)^{-(\frac{1}{\alpha})}\,dq = 1-(1-p)^{1-\frac{1}{\alpha}}.
    \end{equation}
    \item \textit{Discrete wealth distribution: } To obtain the Lorenz function for a discrete income distribution, consider an economy comprising G groups of people, where each group g has $n_g$ individuals with the same income $x_g$ such that $0 \leq x_{1} \leq x_{2} \leq \dots \leq x_{G}$. The total population of the economy is N, and the total income is M, leading to a mean income of $\mu_{g} = M/N$. The income distribution is a discrete random variable X with a probability mass function $f_G(x_g) = n_g/N$ for all g ranging from 1 to G, and a distribution function $F_G(x)$ defined by 0 if $x \in [0, x_1)$, $f_G(x) = n_g/N$ if $x \in [x_g, x_{g+1})$ for any given g ranging from 1 to $(G-1)$, and 1 if $x \geq x_G$.
We define $N(g)$ and $M(g)$ as the cumulative proportion of the population and cumulative proportion of the total income respectively for each group g. For any given g ranging from 1 to G and any $q_g \in (N(g-1), N(g)]$, it can be verified that $F_G^{-1}(q_g) = x_g$. Using the Lorenz function formula, we can calculate the Lorenz function $L_{F_G}(p_g)$ for any given g ranging from $1$ to $G$ and any $p_g \in (N(g-1), N(g)]$ as,
\begin{equation}
   L_{F_G}(p_g) = M(g-1) + (p_g - N(g-1)) \frac{N x_{g}}{M}. 
\end{equation}
We make two observations in this context: 
The first observation states that the Lorenz function is piecewise linear, which means that it is composed of several line segments. The kink points represent the points where the direction of the Lorenz curve changes, and they occur at the boundaries of each income group. At these points, there is a jump in the cumulative share of income that is distributed to each group, which causes a change in the slope of the Lorenz curve.
The second observation is that if there is only one income group in the economy, then the Lorenz curve is a straight line passing through the origin and with a slope of 1. This means that the distribution of income is perfectly equal, and each individual in the economy has the same income. In this case, the Lorenz curve coincides with the diagonal of the unit square, which represents the line $L(p) = p$.
\end{enumerate}

%%%%%%%%%%%%%%%%%%%%%%%%%%%%%%%%%%%%%%

\subsection{\textbf{Properties of Gini index}}
With the above definition it also has several properties that are important to understand when interpreting and using this index.
Some of the key properties of the Gini index are:
\begin{enumerate}
    \item \textit{Range:} The Gini index ranges from 0 to 1, with 0 indicating complete equality (i.e., everyone has the same income or wealth) and 1 indicating complete inequality (i.e., one person has all the income or wealth).

    \item \textit{Normalization:} The Gini index is normalized, meaning that it can be used to compare inequality across different populations or over time. This allows for meaningful comparisons even when the populations or time periods have different sizes or levels of income.

    \item \textit{Sensitivity to changes in the distribution:}  The Gini index is sensitive to changes in the distribution of income or wealth, meaning that even small changes in the distribution can result in large changes in the Gini index. This property makes the Gini index a useful tool for measuring the impact of policies or events that affect the distribution of income or wealth.

    \item \textit{Unimodality:}  The Gini index is unimodal, meaning that it has a single peak. This property allows for the ranking of populations or time periods based on their level of inequality.

    \item \textit{Invariance to scale:} The Gini index is invariant to scale, meaning that it is not affected by changes in the units of measurement (e.g., dollars, euros, etc.). This allows for meaningful comparisons of inequality across populations or time periods that use different currencies.
\end{enumerate}

\subsection{\textbf{Exemplary calculations of Gini index}}
Here we show calculations of Gini index for some simple continuous and discrete wealth distributions.
\begin{enumerate}
    \item \textit{Uniform wealth distribution:} For a uniform distribution on a compact interval $[a,b]$ following $0 \leq a < b < \infty$ leads to the following Gini index,
    \begin{equation}
       g_{u} = 2\int_{0}^{1}\left[q-q\left\{1-\frac{(b-a)}{(a+b)}(1-q)\right\}\right]\,dq = \frac{(b-a)}{3(a+b)}.
    \end{equation}

    \item \textit{Exponential wealth distribution:} An exponential distribution of the form, $F_E(x) = 1-e^{-\lambda x}$ for any $x \geq 0$ and $\lambda > 0$ leads to the following Gini index,
    \begin{equation}
        g_{E} = 2\int_{0}^{1}\left[q-L(q)\right]\,dq = 2\int_{0}^{1}(1-q)\ln{\left(\frac{1}{1-q}\right)}\,dq = \frac{1}{2}.
    \end{equation}
    
    \item \textit{Pareto wealth distribution:} A Pareto distribution of the form $F_{P,\alpha}(x) = 1-(m/x)^{\alpha}$ with $m > 0$ being the minimum income and $ \alpha > 1$ gives a Gini index of the  following form,
    \begin{equation}
        g_{P,\alpha} = 2\int_{0}^{1}\left[q-\left\{1-(1-q)^{1-\frac{1}{\alpha}}\right\}\right]\,dq = \frac{1}{2\alpha -1}.
    \end{equation}
    As we graph the Gini index for various values of $\alpha$ where $\alpha$ is greater than 1, we observe that as $\alpha$ increases, the Gini index decreases. Additionally, as $\alpha$ approaches 1, the Gini index tends towards 1. Furthermore, if we set $\hat{\alpha}$ to be equal to $\frac{\ln{5}}{\ln{4}}$, then the Gini index for $g_{P,\alpha}$ is approximately 0.7565.
    
    \item \textit{Discrete wealth distribution:} Consider the discrete random variable $F_G$ discussed previously for which the Lorenz function is given by Eq.~(12). From this, We have the following explicit form of the Gini index,
    \begin{equation}
        g_{F_{G}} = \frac{\sum_{g=1}^{G}\sum_{t=1}^{G}n_{t}n_{g}|x_{t}-x_{g}|}{2\,N\,M}.
    \end{equation}
    Note that if $n_g = 1$ for all $g \in \{1,...,G \}$ so that $G = N$ and $M = \sum_{g=1}^{N} x_{g}$, then from Eq.~(16) it follows that,
    \begin{equation}
        g_{F_{G}} = \frac{\sum_{g=1}^{G}\sum_{t=1}^{G}|x_{t}-x_{g}|}{2\,N\,\sum_{g=1}^{N}x_{g}}.
    \end{equation}
\end{enumerate}

\subsection{\textbf{Properties of the\texorpdfstring{$k$}{Lg}-index}}
With the above definition $k$-index has several characterizations which are listed below:
\begin{enumerate}
    \item $k$-index as a unique fixed point of the complementary Lorenz function.\\
    \textit{Proof:}
    We can rewrite Eq.~(5) as,
    \begin{equation}
        k=1-L(k)=\Hat{L}(k).
    \end{equation}
Hence, the $k$-index is a fixed point of the complementary Lorenz function. Since the complementary Lorenz function maps [0,1] to [0,1] and is continuous (shown in Fig.~\ref{fig:Lorenz curve}), it has a fixed point by Brouwer's fixed point theorem. Furthermore, since $\Hat{L}(p)$ is non-increasing, the fixed point has to be unique. (proved).

     \item For any distribution, $k \in [1/2,1]$ and the normalized $k$-index, $K=1-2k$ lies in the interval $[0,1]$.\\
     \textit{Proof:}
     Observe that, if $L(p) = p$, then from Eq.~(18),$k = 1/2$ and for any other income distribution, $1/2 < k < 1$. 
     Also note that while the Lorenz curve typically has only two trivial fixed points, that is, $L(0) = 0$ and $L(1) = 1$, the complementary Lorenz function $\Hat{L}(p)$ has a unique non-trivial fixed point $k$.
     Now the normalised $k$-index is given by,$K=1-2k$, so if $k \in [1/2,1]$, then $K \in [0,1]$. (proved).
     
    \item $k$-index as a generalization of the Pareto principle.\\
    \textit{Proof:}
   The k-index can be thought of as a generalization of the Pareto's 80/20 rule. Note that $L(k) = 1 - k$ ; hence, the top $100(1-k)\%$ of the population has $100(1-(1-k)) = 100k\%$
of the income. Hence, the `Pareto ratio' for the k-index is $k/(1 - k)$. Note that this proportion is derived internally from the distribution of income, and typically, there is no expectation that it will align with the Pareto principle. 

    \item Interpreting the $k$-index in terms of rich-poor disparity.\\
    \textit{Proof:} Let's split society into two groups - the `poorest' group, consisting of a fraction p of the population, and the `rich' group, consisting of a fraction (1-p) of the population. Using the Lorenz curve L(p), we can determine the distance between the ``boundary person" and the poorest person on one hand, and the distance between the ``boundary person" and the richest person on the other hand. These distances can be calculated using the equations: $\sqrt{p^{2}+L(p)^{2}}$ and $\sqrt{(1-p)^{2}+(1-L(p))^{2}}$, respectively. The k-index is a way of dividing society into two groups such that the boundary person is equidistant from the poorest and richest persons. The disparity function value at the k-index is given by $D(k) = k - 1/2$. This function measures the gap between the proportion k of the poor from the 50/50 population split. If society is not completely equal, then $k > 1/2$, making it a useful tool to highlight the rich-poor disparity. In this case, k defines the income proportion of the top (1-k) proportion of the rich population.
    
    \item The $k$-index as a solution to optimization problems.\\
    \textit{Proof:}
    The $k$-index is the unique solution to the following surplus maximization problem: 
    \begin{equation}
        k = argmax_{P \in [0,1]}\int_{0}^{P}(\Hat{L}(t)-t)\,dt.
    \end{equation}
    
The value of $k$ is such that it maximizes the area between the complementary Lorenz function and the income distribution line linked with an egalitarian distribution, for the lower-income population. Eq~(19) is a consequence of the fact that $\Hat{L}(p) \geq p$ for all $p \in [0,k]$ and $\Hat{L}(p) \leq p$ for all $p \in (k,1]$. Similarly, the $k$-index is the only solution to the surplus minimization problem (which is the dual of the problem in (19)):
    \begin{equation}
 k = argmin_{P \in [0,1]}\int_{P}^{1}\left\{(1-t)-L(t)\right\}\,dt.
 \end{equation} 
 Therefore, $(1-k)$ is that fraction of the higher income population for which the area between the income distribution line associated with the egalitarian distribution and the Lorenz function is minimized.
 
    \item For reducing inequality between groups, $k$-index is a better indicator.\\
    \textit{Proof:}
      
The ordering of Lorenz curves based on the $k$-index is not the same as the ordering based on the Gini index. While the Gini index is influenced by transfers only within the poor or rich population, the ranking based on the $k$-index is influenced only by transfers between the two groups. This implies that if the objective is to reduce inequality between the groups, then the $k$-index is a more appropriate measure to use.
\end{enumerate}

\subsection{\textbf{Exemplary calculations of \texorpdfstring{$k$}{Lg}-index}}
Here we show calculations of some simple continuous and discrete wealth distribution functions.
Majority part of this subsection is adopted from \citep[][]{k-index2}.
\begin{enumerate}
    \item \textit{Uniform wealth distribution:} If we have the uniform distribution $F$ defined on $[a,b]$ where $0 \leq a < b < \infty$. Then the $k$-index is given by,
    \begin{equation}
        k_{u} = \frac{(3a+b)+\sqrt{5a^{2}+6ab+5b^{2}}}{2(b-a)},
    \end{equation} 
    and the normalized $k$-index is given by,
    \begin{equation}
     K_{u} = \frac{-2(a+b)+\sqrt{5a^{2}+6ab+5b^{2}}}{(b-a)}.
    \end{equation}
    \item \textit{Exponential wealth distribution:} For the exponential distribution $F_{E}$, the complementary Lorenz function is given by $\Hat{L}_{E}(p) = (1 - p) \left[1 + \ln\left({\frac{1}{1-p}}\right)\right]$. One can show that, the $k$-index, $k_{E} \simeq 0.6822$ and hence the normalised $k$-index, $K_{E} \simeq 0.3644$.
    \item \textit{Pareto wealth distribution:} For the Pareto distribution $F_{P,\alpha}$, the complementary Lorenz function is given $L_{P,\alpha}(p) = (1-p)^{1-\frac{1}{\alpha}}$. The $k$-index is therefore a solution to the equation,
    \begin{equation}
      (1-k_{P,\alpha})^{1-\frac{1}{\alpha}} = k_{P,\alpha}.   
    \end{equation}. 
    It is difficult to provide a general solution to this equation. However, we have an interesting observation in this context.
    If $\alpha = \frac{\ln {5}}{\ln {4}} \simeq 1.16$, then the $k$-index, $k_{P,\alpha} = 0.8$ and we get the Pareto principle or the 80/20 rule. Also note that the normalised $k$-index, $K_{P,\alpha} = 0.6$.
    \item \textit{Discrete wealth distribution:} Consider any discrete random variable with the distribution function ($F_{G}$) discussed above for which the Lorenz function is given by \mbox{Equation~(12)}. To~obtain the explicit form of the $k$ index, one can first apply a simple algorithm to identify the interval of the form $[N(g-1),N(g))$ defined for $g \in \{1,\ldots,G\}$ in which the $k$ index can~lie.

Since $N(G)=M(G)=1$, if~we have $N(G-1)+M(G-1)<1$ in some step, then in~the next step, this algorithm has to end, since $N(G)+M(G)=2>1$.

Suppose that for any discrete random variable with the distribution function ($F_G$) discussed earlier, Algorithm-1 identifies $g^*\in \{1,\ldots,G\}$ such that $N(g^*)+M(g^*)\geq 1$. If \mbox{$N(g^*)+M(g^*)=1$}, then $k_{F_G}=N(g^*)$, and if $N(g^*)+M(g^*)>1$, $k_{F_G}$ is the solution to the following equation:
\begin{equation*}
  k_{F_G}+\left\{M(g^*-1)+\left(k_{F_G}-N(g^*-1)\right)\left(\frac{Nx_{g^*}}{M}\right)\right\}=1.
\end{equation*}

Thus, to~derive the $k$ index of any discrete random variable with distribution function $F_G$, we first my identify the group $g^*\in \{1,\ldots,G\}$ such that $k_{F_G}\in (N(g^*-1),N(g^*)]$ (using Algorithm-1); then, using
$g^*$, we obtain the following value of $k_{F_G}$:
\begin{equation}
k_{F_G}= \left\{\begin{array}{cl} N(g^*) & \mbox{ if $N(g^*)+M(g^*)=1$,}\\
 \frac{\mu_G+N(g^*)x_{g^*}-M(g^*)}{\mu_G+x_{g^*}} & \mbox{if $N(g^*)+M(g^*)>1$.}
\end{array} \right.
\end{equation}	
Algorithm-1:
    \begin{enumerate}
\item[Step 1:] Consider the smallest $g_1\in \{1,\ldots,G\}$ such that $N(g_1)\geq 1/2$ and consider the sum of $N(g_1)+M(g_1)$. If~$N(g_1)+M(g_1)\geq 1$, then stop, and $k_{F_G}\in (N_{g_1-1},N(g_1)]$; in~particular, $k_F=N(g_1)$ if and only if $N(g_1)+M(g_1)=1$. Instead, if~$N(g_1)+M(g_1)<1$, then go to Step 2, consider the group $g_1+1$, and repeat the~process. \\
  $\vdots$
	
      \item[Step $t$:] Having reached Step $t$ means that in Step
        $(t-1)$, we had $N(g_1+t-1)+M(g_1+t-1)<1$. Therefore, consider
        the sum of $N(g_1+t)+M(g_1+t)$. If~$N(g_1+t)+M(g_1+t)\geq 1$, then, stop; $k_{F_G}\in [N(g_1+t-1),N(g_1+t))$, and, in~        particular, $k_F=N(g_1+t)$ if and only if
        $N(g_1+t)+M(g_1+t)=1$%Please check intended meaning has been retained
        . If~$N(g_1+t)+M(g_1+t)<1$, then proceed to Step $(t+1)$.
        \end{enumerate}
\end{enumerate}

%%%%%%%%%%%%%%%%%%%%%%%%%%%%%%%%%%%%%%
\section{Some analytical studies on emerging coincidence of Gini and Kolkata indices}\label{sec:4}

\subsection{\textbf{A Landau-like phenomenological expansion of Lorenz function}}
Before going into specific forms of Lorenz curve and their corresponding inequality indices, we will first outline an attempt to have a Landau-like phenomenological expansion of the Lorenz function that obeys the above mentioned properties (see e.g., \citep[][]{bijin}). Particularly, as a minimal non-trivial expansion, we could write
\begin{equation}
    L(p)=Ap+Bp^2,
\end{equation}
with $A>0$, $B>0$ and $A+B=1$. It then follows that calculation of $g=1-2\int\limits_0^1L(p)dp$ and $k=1-L(k)$ give
\begin{equation}
    k=\frac{(3g-2)\pm\sqrt{(2-3g)^2+12g}}{6g},
\end{equation}
which, in the limit $g\to 0$, gives $k=1/2+3g/8$, which implies that if a situation arises when $k=g$, then $k=g=0.8$, which is the Pareto value. 
In the later sections we will discuss the actual situations seen in data and models under unrestricted competitions. 

\subsection{\textbf{Some typical power law forms of Lorenz functions}}
An interesting observation comes out when the Lorenz curve is parameterized through a power law curve in $p$ and for the respective Lorenz curves Gini index and $k$-index are plotted.
When the power law index is made increased, we observe that the values of Gini and $k$-index converges towards each other and meet at a point of around $0.87$ which is not equal to 1.

Here we consider a set of Lorenz functions with the functional form $L(p)=p^{n}$, where $p$ ranges over the interval $[0,1]$, and $n$ is a positive integer greater than or equal to one. 
For this family of functions, we can derive the corresponding Gini index $g_n$, which is equal to $(n-1)/(n+1)$. 
To investigate this relationship further, we have tabulated the Lorenz functions $L(p)$ for $n$ values ranging from 1 to 20, and computed their associated Gini index $g_n$ and $k$-index $k_n$. 
The results of our calculations are presented in Table \ref{tab:1}.

\begin{figure}[H]
\centering
\includegraphics[width=0.65\textwidth]{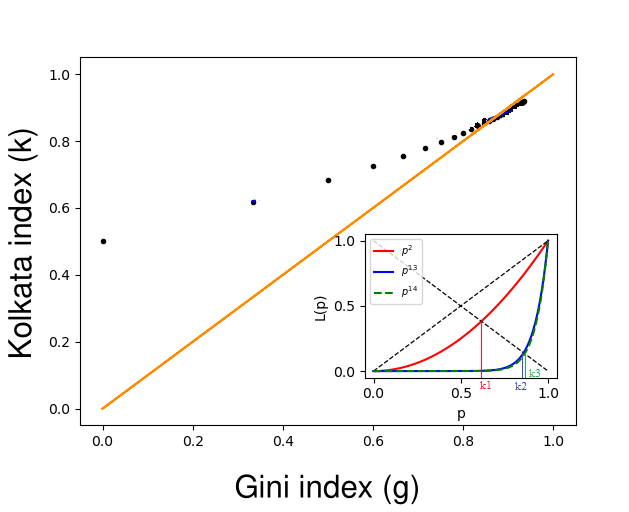}
\caption{Graph of Gini index ($g$) versus $k$-index ($k$), where the orange line represents the equality line ($g=k$). 
The black dots indicate the ($g,k$) values for the Lorenz function, $L(p)=p^{n}$, with $n$ ranging from 1 to 20.
The inset shows the Lorenz curves for $L(p)=p^2$ (red curve), $L(p)=p^{13}$ (blue curve), and $L(p)=p^{14}$ (green dashed curve), with their corresponding $k$-index values ($k1\simeq0.618$, $k2\simeq0.860$, and $k3\simeq0.866$, respectively). This figure is adopted from Banerjee et. al. (2022) \citep[][]{Suchi}.}
\label{fig:lor_2}
\end{figure}

 \begin{table}[H]
 \centering
\begin{tabular}{|c|c|c|c|}
\hline
$n$   &        $L(p)=p^n$  &     $g_{n}$   &        $k_{n}$  \\
\hline
$(n=1)$    &      $p$      &   0           &       $\frac{1}{2}$ \\
\hline
$(n=2)$   &       $p^2$         &   $\frac{1}{3}$          &       $\frac{\sqrt{5}-1}{2}\simeq 0.618$ \\
\hline
$(n=3)$    &      $p^3$      &  $ \frac{1}{2}$   &  0.682\\
\hline
$(n=4)$    &      $p^4$      & $\frac{3}{5}$  &  0.725  \\
\hline
$(n=5)$    &      $p^5$      &   0.667   &          0.755  \\
\hline
$(n=6)$    &      $p^6$      & 0.714    &          0.778  \\
\hline
$(n=7)$    &      $p^7$     &  0.750  &          0.797  \\
\hline
$(n=8)$    &      $p^8$      &  0.778  &        0.812 \\
\hline
$(n=9)$    &      $p^9$     & 0.800       &       0.824\\
\hline
$(n=10)$    &      $p^{10}$      & 0.818  &     0.835  \\
\hline
$(n=11)$    &      $p^{11}$     &  0.833  &  0.844  \\
\hline
$(n=12)$    &      $p^{12}$      & 0.846  &  0.853 \\
\hline
$(n=13)$    &      $p^{13}$      &  0.857  &  0.860 \\
\hline
$(n=14)$    &      $p^{14}$      & 0.867  &  0.866  \\
\hline
$(n=15)$    &      $p^{15}$      & 0.875  &  0.872 \\
\hline
$(n=16)$    &      $p^{16}$      & 0.882  &  0.877 \\
\hline
$(n=17)$    &      $p^{17}$      & 0.889  &  0.882 \\
\hline
$(n=18)$    &      $p^{18}$      &  0.895  &  0.886 \\
\hline
$(n=19)$    &      $p^{19}$      &  0.900  &  0.890  \\
\hline
$(n=20)$    &      $p^{20}$      & 0.905  &  0.894  \\
\hline
\end{tabular}
\caption{The set of Lorenz functions of the form $L(p)=p^{n}$ with $p$ ranging over the interval $[0,1]$ and $n$ values ranging from 1 to 20.}
% The family of Lorenz functions of the form $L(p)=p^{n}$ for all $p\in [0,1]$ where $n\in \{1,\ldots,20\}$.}
\label{tab:1}
\end{table}
In Table \ref{tab:1}, observe that if $n=13$, then $g_{13}\simeq 0.857<k_{13}\simeq 0.860$ and if $n=14$, then $g_{14}\simeq 0.867>k_{14}\simeq 0.866$ implying that coincidence takes place for some $n\in (13,14)$ (see in Fig.~\ref{fig:lor_2}). The coincidence between the Gini index and $k$-index occurs at some $n^*\in (n_1=13.82986,n_2=13.82987)$ and hence we have,   
\begin{equation}
g_{n^*}=k_{n^*}\in \left(g_{n_1}=0.8651369,g_{n_2}=0.8651370\right)\in \left(\frac{6}{7},\frac{13}{15}\right).\hspace{1in}
\end{equation}

Above study shows that there is no positive integer $n$ for which $g$ and $k$ index values coincide and if $n$ represents a positive real number, then there exists a value of $n=n^{*}$ for which these index values coincide.

\subsection{\textbf{Some generic forms of Lorenz functions}}
\begin{figure}[H]
    \centering
    \includegraphics[scale=0.65]{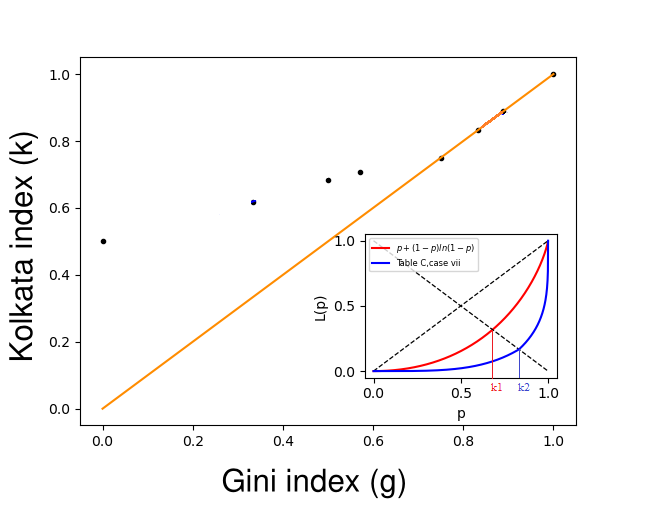}
    \caption{The plot of the Gini index ($g$) versus the $k$-index ($k$), where the orange line corresponds to the $g=k$ line. The black dots represent ($g,k$) values for several simple Lorenz functions, as listed in Table \ref{tab:2}. It is observed that the black dots tend to converge towards the $g=k$ line for larger values of $g$. In the inset, two different Lorenz curves are shown for cases (4) and (7) from Table \ref{tab:2}. The red curve represents the Lorenz curve for case (4) and has a $k$-index value of $k1 \simeq 0.682$, while the blue curve represents the Lorenz curve for case (7) and has a $k$-index value of $k2 \simeq 0.833$. These results provide insight into the relationship between the Gini index and $k$-index, and the behavior of these measures across different Lorenz curves (adopted from \citep[][]{Suchi}).}
    \label{fig:lor_3}
\end{figure}

Therefore, with the above exercise we observe for a specific parameterized Lorenz function where does Gini and $k$-index become equal. 
For more generic Lorenz functions this equality of Gini and $k$-index can be found by the following table \ref{tab:2} (also see Fig.~\ref{fig:lor_3}).

 \begin{table}[H]
 \centering
\begin{tabular}{|c|c|c|c|}
\hline
$Cases$   &        $L_{(n,a)}(p)$  &     $g_{(n,a)}=k_{(n,a)}$   &        ${\rm Interval} \ {\rm of} \ n$  \\
\hline
$(1)$    &      $p^n $      &   0.865           &       (13,14) \\
\hline
$(2)$   &       $\sum\limits_{m=1}^n\left(\frac{1}{n}\right)p^m$         &   0.869          &       (65,66) \\
\hline
$(3)$    &      $\left\{\frac{(n-1)}{n}\right\}p+\sum\limits_{m=2}^{n}\left\{\frac{1}{n(n-1)}\right\}p^m $      &  $ \rm Impossibility$   &  $--$\\
\hline
$(4)$    &      $\sum\limits_{m=1}^{n-1}\left\{\frac{1}{n(n-1)}\right\}p^m + \left\{\frac{(n-1)}{n}\right\}p^{n} $      &   0.874  &  (17,18)  \\
\hline
$(5)$    &      $\sum\limits_{m=1}^{n}\left\{\frac{6m(n+1-m)}{n(n+1)(n+2)}\right\}p^m$      &   0.874   &         (40,41)  \\
\hline
$(6)$    &      $\sum\limits_{m=1}^{n}\left\{\frac{2(m+1)}{n(n+3)}\right\}p^m$       & 0.877    &        (29,30)  \\
\hline
$(7)$    &      $\sum\limits_{m=1}^{n}\left\{\frac{\ln(n+1-m)}{\sum\limits_{r=1}^n\ln(n+1-r)}\right\}p^m $     &  0.881  &          (77,78)  \\
\hline
\end{tabular}
\caption{Different polynomial Lorenz functions (adopted from Banerjee et.al. \citep[][]{Suchi}).}
\label{tab:2}
\end{table}
In this review article, we examine the properties of finite polynomial Lorenz functions through seven cases presented in Table \ref{tab:2}. The first case (1) has already been addressed, where $a_1=\ldots=a_{n-1}=0$ and $a_n=1$. In Case (2), $a_1=\ldots=a_n=1/n$, which results in a coincidence value of approximately 0.869 for some value between 65 and 66. This coincidence value is higher than that obtained under Case (1). In Case (3), $a_1=(n-1)/n$ and $a_2=\ldots=a_n=1/[n(n-1)]$. Since the weight $a_1$ is relatively high, $k$ is close to 1/2, and thus there is no coincidence between the Gini index and the $k$ index. In Case (4), $a_n=(n-1)/n$ and $a_1=\ldots=a_{n-1}=1/[n(n-1)]$, resulting in a coincidence value of approximately 0.874 for some value between 17 and 18, which is higher than that under Case (2). In Case (5), $a_m=[6m(n+1-m)]/[n(n+1)(n+2)]$ for all $m={1,2,\ldots,n}$, and the coincidence value is 0.874 for some number lying between 40 and 41, which is not an improvement compared to Case (4). In Case (6), $a_m=[2(m+1)]/[n(n+3)]$ for all $m={1,2,\ldots,n}$, and the coincidence value is 0.877 for some value between 29 and 30. Finally, in Case (7), the maximum coincidence value of 0.881 is achieved for some value between 77 and 78, where, $a_m=\sum_{m=1}^n \left[ \ln (n+1-m)/\left\{\sum_{r=1}^n\ln(n+1-r)\right\}\right]$ for all $m={1,2,\ldots,n}$. Overall, our analysis shows that the coincidence value is less than 8/9 (approximately 0.88) for all seven cases discussed in Table \ref{tab:2}.
 
Coincidentally with the data from different scenarios it has been found that the inequality in those scenarios reach a maximum point of around $0.87$ before dropping. 

\section{Real-world data indicating the convergence of the Gini and Kolkata indices in various socio-economic contexts}\label{sec:5}

In this section, we investigate the emergence of pervasive inequality as an observable example of emergent properties in different socio-economic systems. 
These systems exhibit universal characteristics of the inequality indices, Gini ($g$) and Kolkata ($k$), as shown by real data collected from diverse social and economic systems.
Our data collection and analysis were completed before the end of 2021. 
It is evident that these systems show emergent properties when their dynamics are not externally fine-tuned.
In the current situation, this leads to an environment of unrestricted competition. 
To illustrate this phenomenon, we consider a range of socio-economic systems, including income, income tax data, box office earnings from Hollywood (US) and Bollywood (India), Bitcoin price fluctuations based on daily prices, election candidates (vote shares), universities (excellence/citation sharing), publications by authors (citation shares), wars or social conflicts (human death shares), sports (Olympic medals share), among others. These systems demonstrate the emergence of significant and widespread inequality indices, which we explore in detail in this section.

\subsection{\textbf{Socio-Economic Disparities: An Analysis of Income, Income Tax, and Box Office Earnings Data}}\label{subsec:1}
\begin{figure}[H]
\centering
\includegraphics[width=0.65\textwidth]{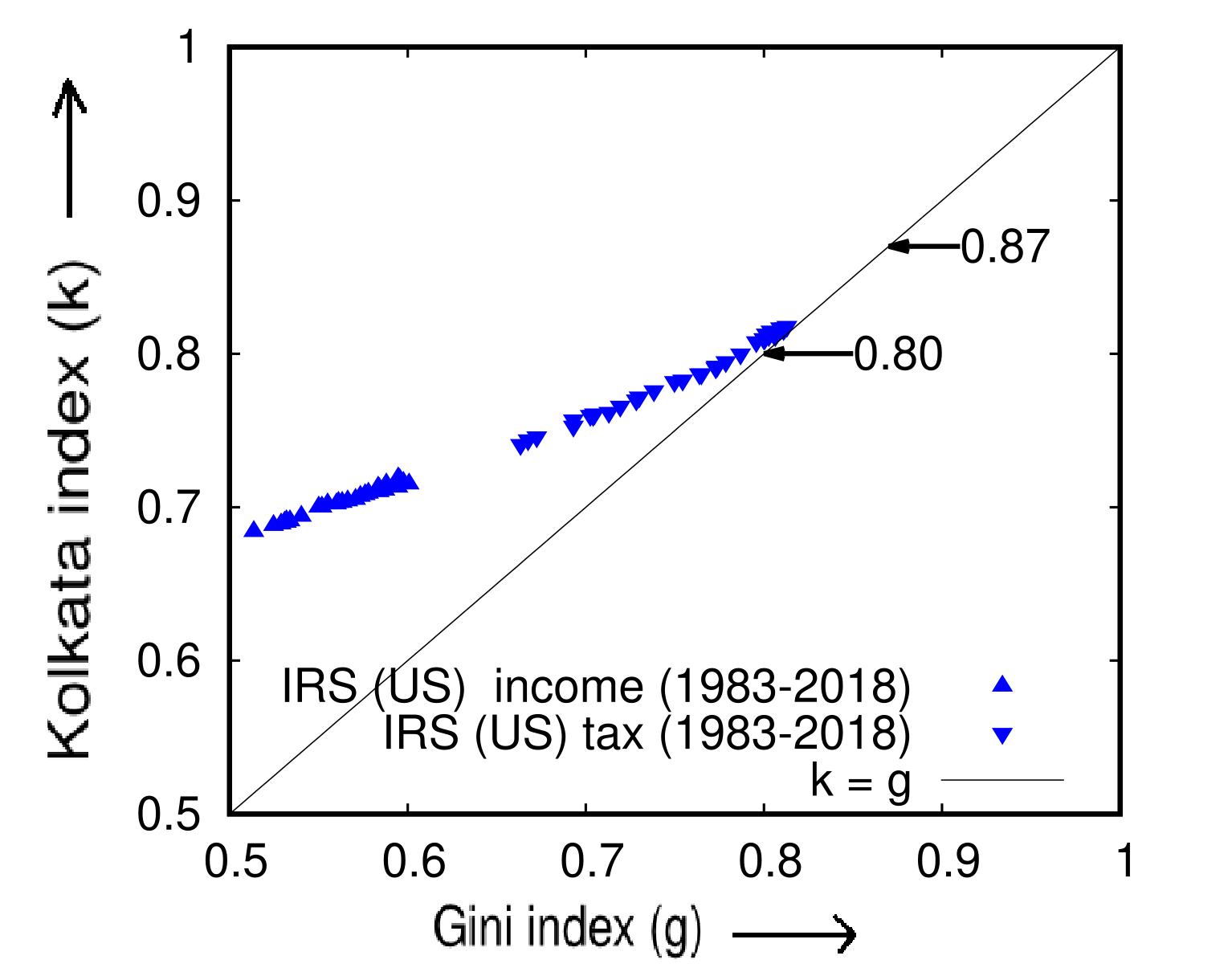}
\caption{Plot of Kolkata index ($k$) against Gini index ($g$) for income and income tax data extracted from IRS (USA) data \citep[][]{IRS},\citep[][]{Ludwig2021} from the years 1983 to 2018. The data was obtained from the corresponding Lorenz functions $L(p)$ for each of these 36 years (adopted from \citep[][]{Suchi}).}
\label{fig:5a}
\end{figure}

In the report by the United Nations Development Program \citep[][]{UNDP1992}, it was found that the global distribution of income is highly unequal, with the top $20\%$ of the world's population receiving $82.7\%$ of the world's income. 
Despite concerns that the top $1\%$ of wealthy individuals possess $99\%$ of the world's wealth (as in `Occupy Wall Street' protest), this observation suggests that $82\%$ of the world's wealth is actually owned by $18\%$ of the population.
To further explore this issue, we analyzed data from the IRS (US) \citep[][]{IRS} and \citep[][]{Ludwig2021} on cumulative income by the poorest $p$ fraction of people over a period of 36 years (1983-2018) and calculated the Lorenz function $L(p)$. We also computed the Gini ($g$) and Kolkata ($k$) indices for each year (see Fig.\ref{fig:5a}).

\begin{figure}[H]
\centering
\includegraphics[width=0.7\textwidth]{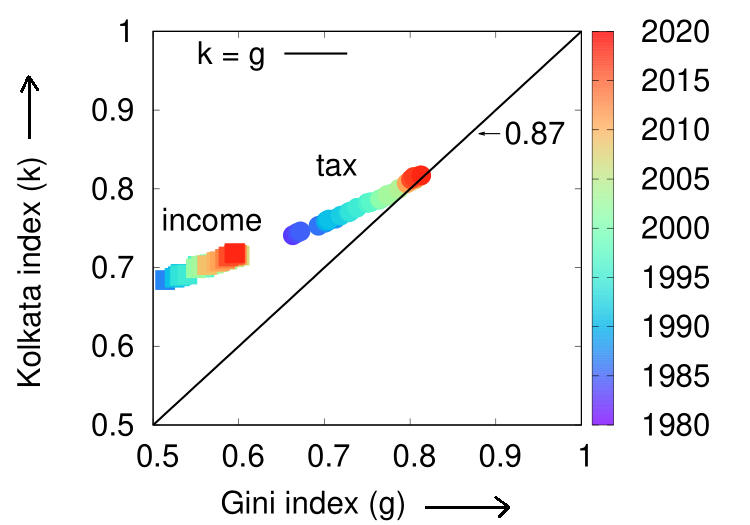}
\caption{Trend of Gini ($g$) and Kolkata ($k$) indices over time (year) for the US economy using IRS data \citep[][]{IRS, Ludwig2021}. The graph clearly shows an increasing trend in the inequality measures over time, indicating a decline in public welfare and a shift towards an SOC state of unrestricted competition. The value of $k$ in the tax data, which is argued to be a better indicator of the prevailing inequality status, surpasses the Pareto value of 0.80 and is predicted to reach 0.87, similar to other socio-economic systems (e.g., movie income or citations), where public welfare programs are completely absent (this figure is adopted from \citep[][]{Suchi}).}
\label{fig:X}
\end{figure}
We present a visual representation of the analysis conducted on the IRS income and tax data \citep[][]{IRS, Ludwig2021} for a period of 36 years (1983-2018) in Figure \ref{fig:X}. The graph depicts a consistent increase in the inequality indices $g$ and $k$ over the years, which converge towards a value of 0.87.

\begin{figure}[H]
\centering
\includegraphics[width=0.7\textwidth]{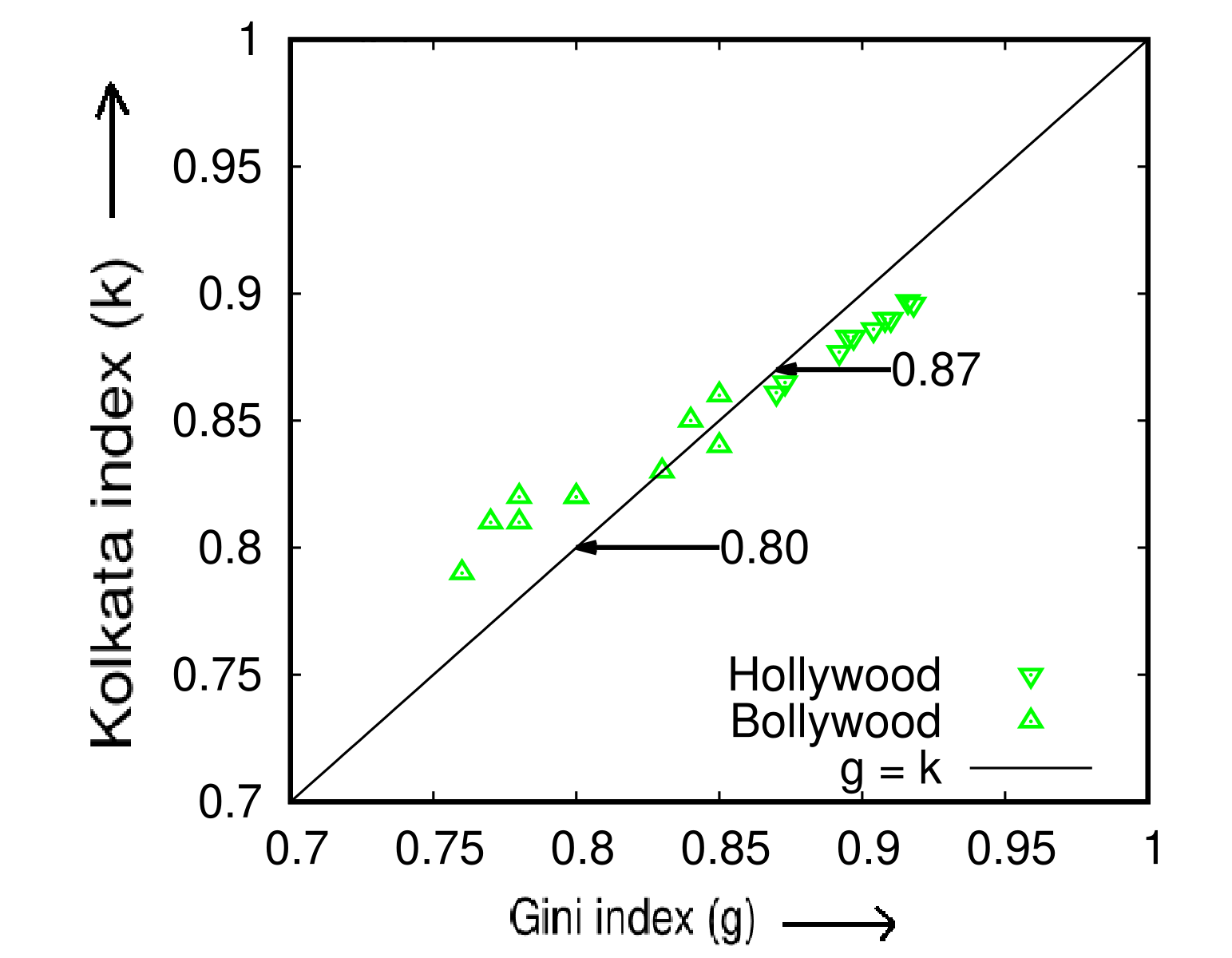}
\caption{A scatter plot of the Kolkata index ($k$) versus the Gini index ($g$) for box office income obtained from Hollywood (USA, data source: \citep[][]{Hollywood2011}) and Bollywood (India, data source: \citep[][]{Bollywood2011}) over a period of 9 years, from 2011 to 2019. The plot provides a comparative analysis of the inequality measures for these two major film industries (adopted from \citep[][]{Suchi}).}
\label{fig:5b}
\end{figure}

Similarly, we extend our analysis to the yearly income generated by the film industry in Hollywood (USA \citep[][]{Hollywood2011}) and Bollywood (India \citep[][]{Bollywood2011}) for a period of 9 years (2011-2019). Table \ref{tab:h_b} and Figure \ref{fig:5b} demonstrate the results of our analysis on the income data for these two different film industries and show that both in Hollywood and in Bollywood, the box-office income inequality index $k$ increases on average to $0.88$ and $0.83$ respectively.

\begin{table}[H]
\centering
\begin{tabular}{|c|c|c|c|}
\hline
 movie & \multicolumn{3}{c|}{\shortstack[lb]{Box office collection \\ from Hollywood (USA) movies}}\\
 \cline{2-4}
 releasing & Total & Gini & Kolkata \\
 year & movies & ($g$) & ($k$)  \\
\hline
2010&651&0.87&0.86\\
\hline
2011&730&0.87&0.87 \\
\hline
2012&807&0.89&0.88 \\
\hline 
2013&826&0.90&0.88 \\
\hline
2014&849&0.90&0.88 \\
\hline
2015&847&0.91&0.89 \\
\hline
2016&856&0.90&0.89 \\
\hline
2017&852&0.91&0.89\\
\hline
2018&993&0.92&0.90 \\
\hline
2019&911&0.92&0.90 \\
\hline
\end{tabular}
\begin{tabular}{|c|c|c|c|}
 \hline
 movie & \multicolumn{3}{c|}{\shortstack[lb]{Box office collection \\ from Bollywood (India) movies}}\\
 \cline{2-4}
releasing & Total & Gini & Kolkata\\
year & movies & ($g$) & ($k$)  \\
\hline
2010&139&0.77&0.81\\
\hline
2011&123&0.78&0.82\\
\hline
2012&132&0.78&0.81\\
\hline
2013&136&0.76&0.79\\
\hline
2014&145&0.8&0.82\\
\hline
2015&166&0.8&0.82\\
\hline
2016&215&0.83&0.83\\
\hline
2017&251&0.85&0.84\\
\hline
2018&218&0.84&0.85\\
\hline
2019&246&0.85&0.86\\
\hline
\end{tabular}
\caption{An analysis of income inequality in box-office earnings for movies released during the period of 2011-2019 in two major film industries, Hollywood (USA) and Bollywood (India). Data taken from refs.\citep[][]{Hollywood2011,Bollywood2011}.}
\label{tab:h_b} 
\end{table}

In this case we will show that almost $88\%$ of Box-Office income share comes only from $12\%$ of Hollywood movies measured from 2011 to 2019. Similarly it is also shown that almost $83\%$ of Box-Office income share comes only from $17\%$ of Bollywood movies measured from 2011 to 2019.

\subsection{\textbf{Inequality in Bitcoin Value Fluctuations: A Data Analysis {Study}}}\label{subsec:2}
Bitcoin is a decentralized cryptocurrency that operates on a ledger system without any central bank control. Its introduction in 2008 and adoption in 2009 have made it the first and largest cryptocurrency, with a market value surpassing $1.03$ as of November 2021, accounting for $2.9\%$ of the world's total narrow money supply. While its decentralized nature has made it susceptible to market volatility, Bitcoin serves as an example of unrestricted competition in currency markets.

\begin{figure}[H]
\centering
\includegraphics[width=0.6\textwidth]{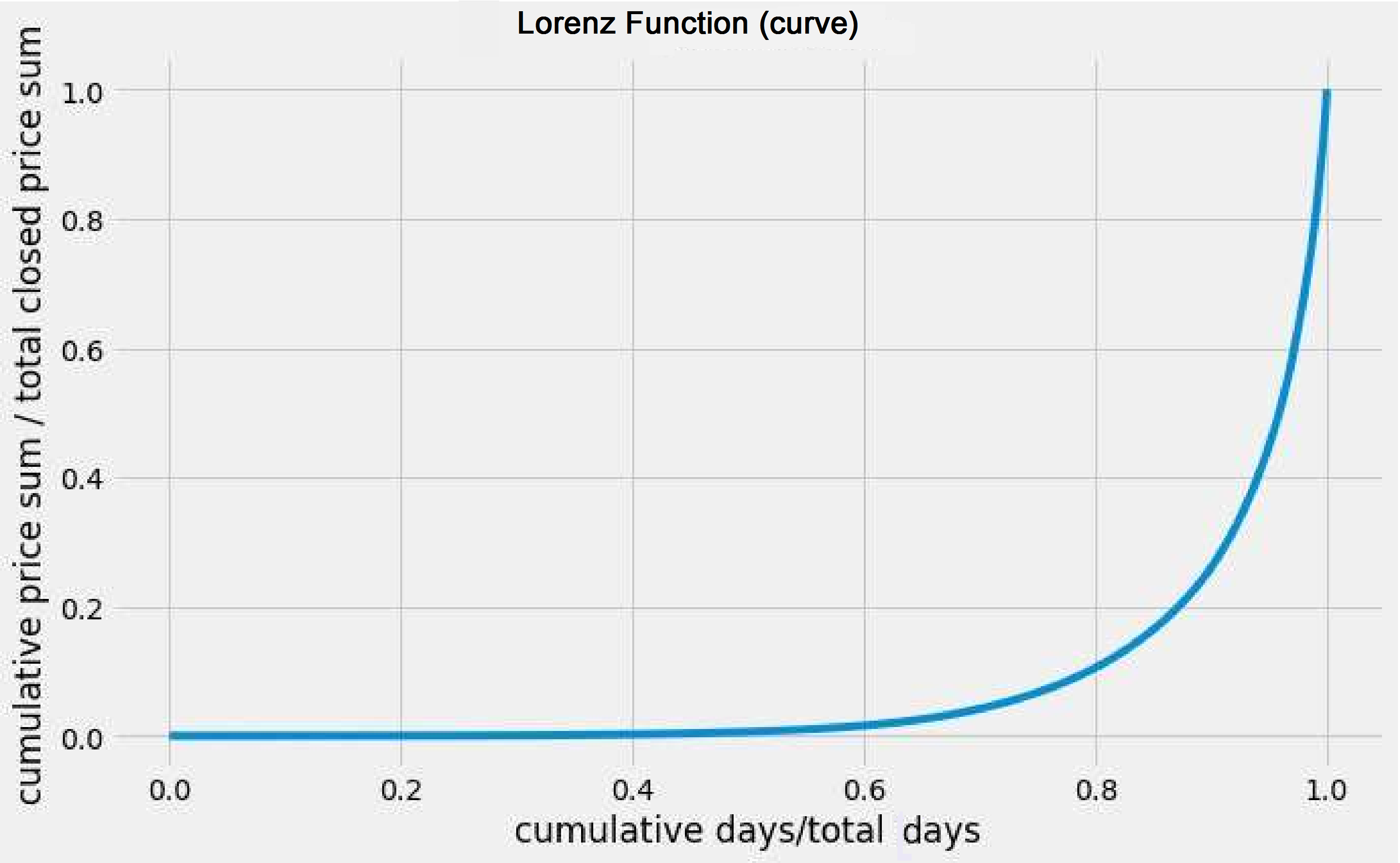}
 \caption{The Lorenz Function (curve) depicts the distribution of the difference in the closing price of Bitcoins for consecutive days (adopted from \citep[][]{Suchi}).}
\label{fig:6a}
\end{figure}

\begin{figure}[H]
\centering
\includegraphics[width=0.95\textwidth]{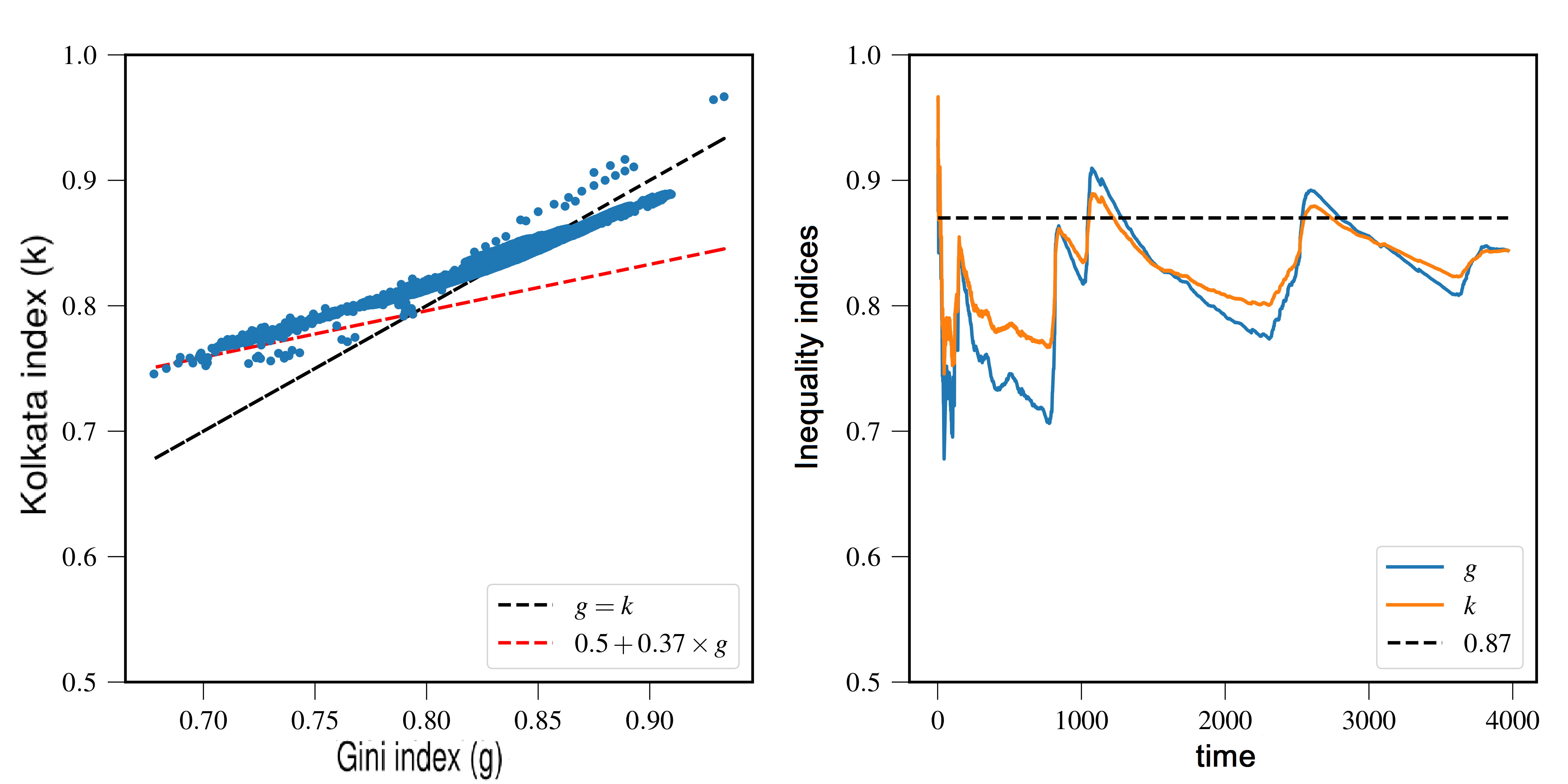}
 \caption{Left: A graphical representation of the Kolkata index ($k$) plotted against the Gini index ($g$) for the statistical analysis of the daily Bitcoin prices.
Right: The temporal variation of the $g$ and $k$ indices are illustrated. For comparison, a reference value of approximately 0.87 is provided in the figure (adopted from \citep[][]{Suchi}).}
\label{fig:6b}
\end{figure}

In this study, we analyze the fluctuation of Bitcoin's value, which is measured in USD, by collecting daily price data from January 1, 2010, to November 24, 2021, using data obtained from \citep[][]{Bitcoin2021}. To investigate its value fluctuation, we calculate the absolute value of the price changes in consecutive days and collect the fractional closing price changes, except for a few days up to a date $t$ $(>t_0)$. We then generate the Lorenz curve (refer to Fig.\ref{fig:6a}) for the closing price data up to a time $t$ $(>>t_0)$ and proceed to estimate the $g$ and $k$ indices as discussed in Fig.\ref{fig:6b}.

The figure (Fig. \ref{fig:6b}) demonstrates that $g$ and $k$ exhibit a pattern of repeatedly approaching each other near the value of $0.87$.
%As can be seen from Fig. \ref{fig:6b}, the values of $g$ and $k$ repeatedly approach each other near 0.87.

\subsection{\textbf{Inequality analysis of vote data for election contestants}}\label{subsec:3}

This subsection analyzes the inequality of vote shares among the candidates in the Indian parliamentary elections held in 2014 and 2019. Table \ref{tab:3} demonstrates that there exists a high level of inequality in the distribution of vote shares. The Gini and Kolkata indices were found to be 0.83 and 0.86 for the 2014 election, and 0.85 and 0.88 for the 2019 election, respectively. The values of these indices are similar to the value of approximately 0.87 observed in the case of unrestricted competition.

 \begin{table}[H]
 \centering
\begin{tabular}{|c|c|c|c|}
\hline
Year   &       Total voters  &      $g$   &        $k$  \\
\hline
2014    &      $5\times10^8$      &   0.83           &       0.86 \\
............&....................&............&............\\
2019   &       $6\times10^8$         &  0.85          &       0.88 \\
\hline
\end{tabular}
\caption{Gini ($g$) and Kolkata ($k$) index values obtained from the Lorenz function $L(p)$ for the vote shares in the Indian Parliament elections held in the years 2014 and 2019, where the number of contesting candidates was over 8000 in each year. The data used for the analysis is obtained from references \citep[][]{Lokesabha2014,Lokesabha2019}.}
\label{tab:3}
\end{table}

\subsection{\textbf{Inequality analysis for citations data of different Journals and Universities}}\label{subsec:4}
Data obtained from ISI web of science \citep[][]{Chatterjee2016} showed that, the citations of the papers published from different leading universities or institutions and leading journals are also unequal and here we take the average value from 1980 to 2010 (see Tables \ref{tab:4},\ref{tab:5} and \ref{tab:6}). 
From the following table (in Table ~\ref{table}), we can easily say that the successful $18-25\%$ of the papers published from different universities or institutes and journals received $82-76\%$ of the citations.

\begin{table}[H]
 \centering
\begin{tabular}{|c|c|c|c|}
\hline
 Inst./Univ./Journ & Papers(\%) & Citations(\%) & Comments\\
 \hline
 Harvard (Univ) & 22 & 78 & About 23\% of the papers \\
\cline{1-3}
MIT (Univ) & 22 & 78 & published from leading \\
\cline{1-3}
IISC (Inst) & 25 & 75 & Univ/Inst received 77\%  \\
\cline{1-3}
TIFR (Inst) & 23 & 77 & of the citations. \\
 \hline 
 & & & About 19\% of the papers \\
 \cline{1-3}
Nature (Journ) & 18 & 82 & published in leading \\
\cline{1-3}
Science (Journ) & 19 & 81 & Journals received 81\% of  \\
\cline{1-3}
& & & the citations.\\
\hline 
 \end{tabular}
 \caption{Average citation share of the papers published (during 1980 to 2010) from different Univ/Inst and in Journals \citep[][]{Chatterjee2016}.}
    \label{table}
\end{table}

% \begin{figure}[H]
%     \centering
%     \includegraphics[scale=0.9]{table.png}
%     \caption{}
%     \label{fig:8}
% \end{figure}

\begin{table}[H]
 \centering
\begin{tabular}{|c|c|c|c|c|c|}
\hline
 Inst./Univ. & Year & \multicolumn{4}{c|}{ISI web of science data}\\
 \cline{3-6}
 &  & $N_p$ & $N_c$  & \multicolumn{2}{c|}{index values}\\
 \cline{5-6}
 & &  & & $g$ &$k$\\
 \hline
 \multirow{4}{*}{\shortstack[lb]{ Melbourne }}  & 
 1980 & 866 & 16107 & 0.67 & 0.75 \\
\cline{2-6}
& 1990 & 1131 & 30349 & 0.68 & 0.75 \\
\cline{2-6}
& 2000 & 2116 & 57871 & 0.65 & 0.74 \\
\cline{2-6}
& 2010 & 5255 & 63151 & 0.68 & 0.75 \\
 \hline 
 \multirow{4}{*}{\shortstack[lb]{ Tokyo }} & 
 1980 & 2871 & 60682 & 0.69 & 0.76 \\
 \cline{2-6}
& 1990 & 4196 & 108127 & 0.68 & 0.76 \\
\cline{2-6}
& 2000 & 7955 & 221323 &  0.70 & 0.76 \\
\cline{2-6}
& 2010 & 9154 & 91349 & 0.70 & 0.76 \\
\hline 
 \multirow{4}{*}{\shortstack[lb]{ Harvard }} & 
1980 & 4897 & 225626 & 0.73 & 0.78 \\
\cline{2-6}
 & 1990 & 6036 & 387244 & 0.73 & 0.78 \\
 \cline{2-6}
 & 2000 & 9566 & 571666 & 0.71 & 0.77 \\
 \cline{2-6}
& 2010 & 15079 & 263600 & 0.69 & 0.76 \\
\hline  
 \end{tabular}
 \begin{tabular}{|c|c|c|c|c|c|}
 \hline
 Inst./Univ. & Year & \multicolumn{4}{c|}{ISI web of science data}\\
 \cline{3-6}
  &  & $N_p$ & $N_c$  & \multicolumn{2}{c|}{index values}\\
 \cline{5-6}
 & &  & & $g$ & $k$\\
 \hline 
  \multirow{4}{*}{\shortstack[lb]{ MIT }} & 
 1980 & 2414 & 101929 & 0.76 & 0.79 \\
 \cline{2-6}
& 1990 & 2873 & 156707 & 0.73 & 0.78 \\
\cline{2-6}
& 2000 & 3532 & 206165 & 0.74 & 0.78 \\
\cline{2-6}
& 2010 & 5470 & 109995 & 0.69 & 0.76 \\
\hline 
\multirow{4}{*}{\shortstack[lb]{ Cambridge }} & 
1980 & 1678 & 62981 & 0.74 & 0.78 \\
\cline{2-6}
& 1990 & 2616 & 111818 & 0.74 & 0.78 \\
\cline{2-6}
& 2000 & 4899 & 196250 & 0.71 & 0.77 \\
\cline{2-6}
& 2010 & 6443 & 108864 & 0.70 & 0.76 \\
\hline
\multirow{4}{*}{\shortstack[lb]{ Oxford }} & 
1980 & 1241 & 39392 & 0.70 & 0.77 \\
\cline{2-6}
& 1990 & 2147 & 83937 & 0.73 & 0.78 \\
\cline{2-6}
& 2000 & 4073 & 191096 & 0.72 & 0.77 \\
\cline{2-6}
& 2010 & 6863 & 114657 & 0.71 & 0.76 \\
\hline 
 \end{tabular}
 \caption{Gini and Kolkata indices for citation
inequalities of publications by authors from
some selected  universities, analyzed in
December 2013,  using the data obtained from
ISI web of science. (Adapted from\citep[][]{k-index},\citep[][]{CG2017}).}
\label{tab:4}
\end{table}

\begin{table}[H]
 \centering
\begin{tabular}{|c|c|c|c|c|c|}
\hline
 Inst./Univ. & Year & \multicolumn{4}{c|}{ISI web of science data}\\
 \cline{3-6}
 &  & $N_p$ & $N_c$  & \multicolumn{2}{c|}{index values}\\
 \cline{5-6}
 & &  & & $g$ &$k$\\
 \hline
 \multirow{4}{*}{\shortstack[lb]{ SINP }}  & 
 1980 & 32 & 170 & 0.72 & 0.74 \\
 \cline{2-6}
& 1990 & 91 & 666 & 0.66 & 0.73 \\
\cline{2-6}  		
& 2000 & 148 & 2225 & 0.77 & 0.79 \\
\cline{2-6}  		
& 2010 & 238 & 1896 & 0.71 & 0.76 \\
\hline
 \multirow{4}{*}{\shortstack[lb]{ IISC }} & 
 1980 & 450 & 4728 & 0.73 & 0.78 \\
 \cline{2-6}
& 1990 & 573 & 8410 & 0.70 & 0.76 \\
\cline{2-6}
& 2000 & 874 & 19 167 &  0.67 & 0.75 \\
\cline{2-6}
& 2010 & 1624 & 11 497 & 0.62 & 0.73 \\
\hline 
 \multirow{4}{*}{\shortstack[lb]{ TIFR }} & 
 1980 & 167 & 2024 & 0.70 & 0.76 \\
 \cline{2-6}
& 1990 & 303 & 4961 & 0.73 & 0.77 \\
\cline{2-6}
& 2000 & 439 & 11 275 & 0.74 & 0.77 \\
\cline{2-6}
& 2010 & 573 & 9988 & 0.78 & 0.79 \\
\hline  
 \end{tabular}
 \begin{tabular}{|c|c|c|c|c|c|}
 \hline
 Inst./Univ. & Year & \multicolumn{4}{c|}{ISI web of science data}\\
 \cline{3-6}
  &  & $N_p$ & $N_c$  & \multicolumn{2}{c|}{index values}\\
 \cline{5-6}
 & &  & & $g$ & $k$\\
 \hline 
  \multirow{4}{*}{\shortstack[lb]{ Calcutta }} & 
  1980 & 162 & 749 & 0.74 & 0.78 \\
  \cline{2-6}
  & 1990 & 217 & 1511 & 0.64 & 0.74 \\
  \cline{2-6}
  & 2000 & 173 & 2073 & 0.68 & 0.74 \\
  \cline{2-6}
  & 2010 & 432 & 2470 & 0.61 & 0.73 \\
  \hline 
\multirow{4}{*}{\shortstack[lb]{ Delhi }} & 
1980 & 426 & 2614 & 0.67 & 0.75 \\
\cline{2-6}
& 1990 & 247 & 2252 & 0.68 & 0.76 \\
\cline{2-6}
& 2000 & 301 & 3791 & 0.68 & 0.76 \\
\cline{2-6}
& 2010 & 914 & 6896 & 0.66 & 0.74 \\
\hline 
\multirow{4}{*}{\shortstack[lb]{ Madras }} & 
1980 & 193 & 1317 & 0.69 & 0.76 \\
\cline{2-6}
& 1990 & 158 & 1044 & 0.68 & 0.76 \\
\cline{2-6}
& 2000 & 188 & 2177 & 0.64 & 0.73 \\
\cline{2-6}
& 2010 & 348 & 2268 & 0.78 & 0.79 \\
\hline 
\end{tabular}
\vskip1ex
\caption{Gini and Kolkata indices for citation
inequalities of publications by authors from
some selected Indian universities/Institutes,
analyzed in December 2013,  using the data
obtained from ISI web of science. (Adapted from 
\citep[][]{k-index}).}
\label{tab:5}
\end{table}

\begin{table}[H]
 \centering
\begin{tabular}{|c|c|c|c|c|c|}
\hline
 Inst./Univ. & Year & \multicolumn{4}{c|}{ISI web of science data}\\
 \cline{3-6}
 &  & $N_p$ & $N_c$  & \multicolumn{2}{c|}{index values}\\
 \cline{5-6}
 & &  & & $g$ &$k$\\
 \hline
 \multirow{4}{*}{\shortstack[lb]{ Nature  }}  & 
 1980 & 2904 & 178927 & 0.80 & 0.81 \\
 \cline{2-6} 	
& 1990 & 3676 & 307545 & 0.86 & 0.85 \\
 \cline{2-6} 	
& 2000 & 3021 & 393521 & 0.81 & 0.82 \\
 \cline{2-6} 	
& 2010 & 2577 & 100808 & 0.79 & 0.81\\
\hline
 \multirow{4}{*}{\shortstack[lb]{ Science }} & 
 1980 & 1722 & 111737 & 0.77 & 0.80 \\
  \cline{2-6} 
  & 1990 & 2449 & 228121 & 0.84 & 0.84 \\
  \cline{2-6} 
  & 2000 & 2590 & 301093 & 0.81 & 0.82\\
   \cline{2-6} 
   & 2010 & 2439 & 85879 & 0.76 & 0.79 \\
\hline 
 \multirow{4}{*}{\shortstack[lb]{ PNAS(USA) }} & 
 1980 & - & - & - & - \\
 \cline{2-6} 
& 1990 & 2133 & 282930 & 0.54 & 0.70 \\
\cline{2-6} 
& 2000 & 2698 & 315684 & 0.49 & 0.68 \\
\cline{2-6} 
& 2010 & 4218 & 116037 & 0.46 & 0.66 \\
\hline  
\multirow{4}{*}{\shortstack[lb]{ Cell }} & 
 1980 & 394 & 72676 & 0.54 & 0.70 \\
 \cline{2-6}
& 1990 & 516 & 169868 & 0.50 & 0.68 \\
\cline{2-6}
& 2000 & 351 & 110602 & 0.56 & 0.70 \\
\cline{2-6}
& 2010 & 573 & 32485 & 0.68 & 0.75 \\
\hline
\multirow{4}{*}{\shortstack[lb]{ PRL }} & 
1980 & 1196 & 87773 & 0.66 & 0.74 \\
\cline{2-6}
& 1990 & 1904 & 156722 & 0.63 & 0.74 \\
\cline{2-6}
& 2000 & 3124 & 225591 & 0.59 & 0.72 \\
\cline{2-6}
& 2010 & 3350 & 73917 & 0.51 & 0.68 \\
\hline
 \end{tabular}
 \begin{tabular}{|c|c|c|c|c|c|}
 \hline
 Inst./Univ. & Year & \multicolumn{4}{c|}{ISI web of science data}\\
 \cline{3-6}
  &  & $N_p$ & $N_c$  & \multicolumn{2}{c|}{index values}\\
 \cline{5-6}
 & &  & & $g$ & $k$\\
 \hline 
\multirow{4}{*}{\shortstack[lb]{ PRA }} & 
1980 & 639 & 24802 & 0.61 & 0.73 \\
\cline{2-6}
& 1990 & 1922 & 54511 & 0.61 & 0.72 \\
\cline{2-6}
& 2000 & 1410 & 38948 & 0.60 & 0.72 \\
\cline{2-6}
& 2010 & 2934 & 26314 & 0.53 & 0.69 \\
\hline 
\multirow{4}{*}{\shortstack[lb]{ PRB }} & 
1980 & 1413 & 62741 & 0.65 & 0.74 \\
\cline{2-6}
& 1990 & 3488 & 153 521 & 0.65 & 0.74 \\
 \cline{2-6}
 & 2000 & 4814 & 155172 & 0.59 & 0.72 \\
 \cline{2-6}
 & 2010 & 6207 & 70612 & 0.53 & 0.69 \\
\hline
\multirow{4}{*}{\shortstack[lb]{ PRC }} & 
1980 & 630 & 19373 & 0.66 & 0.75 \\
\cline{2-6}
& 1990 & 728 & 15312 & 0.63 & 0.73 \\
\cline{2-6}
& 2000 & 856 & 19143 & 0.57 & 0.71 \\
\cline{2-6}
& 2010 & 1061 & 11764 & 0.56 & 0.70 \\
\hline  
\multirow{4}{*}{\shortstack[lb]{ PRD }} & 
1980 & 800 & 36263 & 0.76 & 0.80 \\
\cline{2-6}
& 1990 & 1049 & 33257 & 0.68 & 0.76 \\
\cline{2-6}
& 2000 & 2061 & 66408 & 0.61 & 0.73 \\
\cline{2-6}
& 2010 & 3012 & 40167 & 0.54 & 0.69 \\
\hline  
\multirow{4}{*}{\shortstack[lb]{ PRE }} & 
1980 & - & - & - & -\\
\cline{2-6}
& 1990 & - & - & - & - \\
\cline{2-6}
& 2000 & 2078 & 51860 & 0.58 & 0.71 \\
\cline{2-6}
& 2010 & 2381 & 16605 & 0.50 & 0.68 \\
\hline  
\end{tabular}
\vskip1ex
\caption{Gini and Kolkata indices for citation
inequalities of publications in some selected
science journals in December 2013,  using the
data obtained from ISI web of science. (Adapted from \citep[][]{k-index}).}
\label{tab:6}
\end{table}

\begin{figure}[H]
\centering
\includegraphics[width=0.7\textwidth]{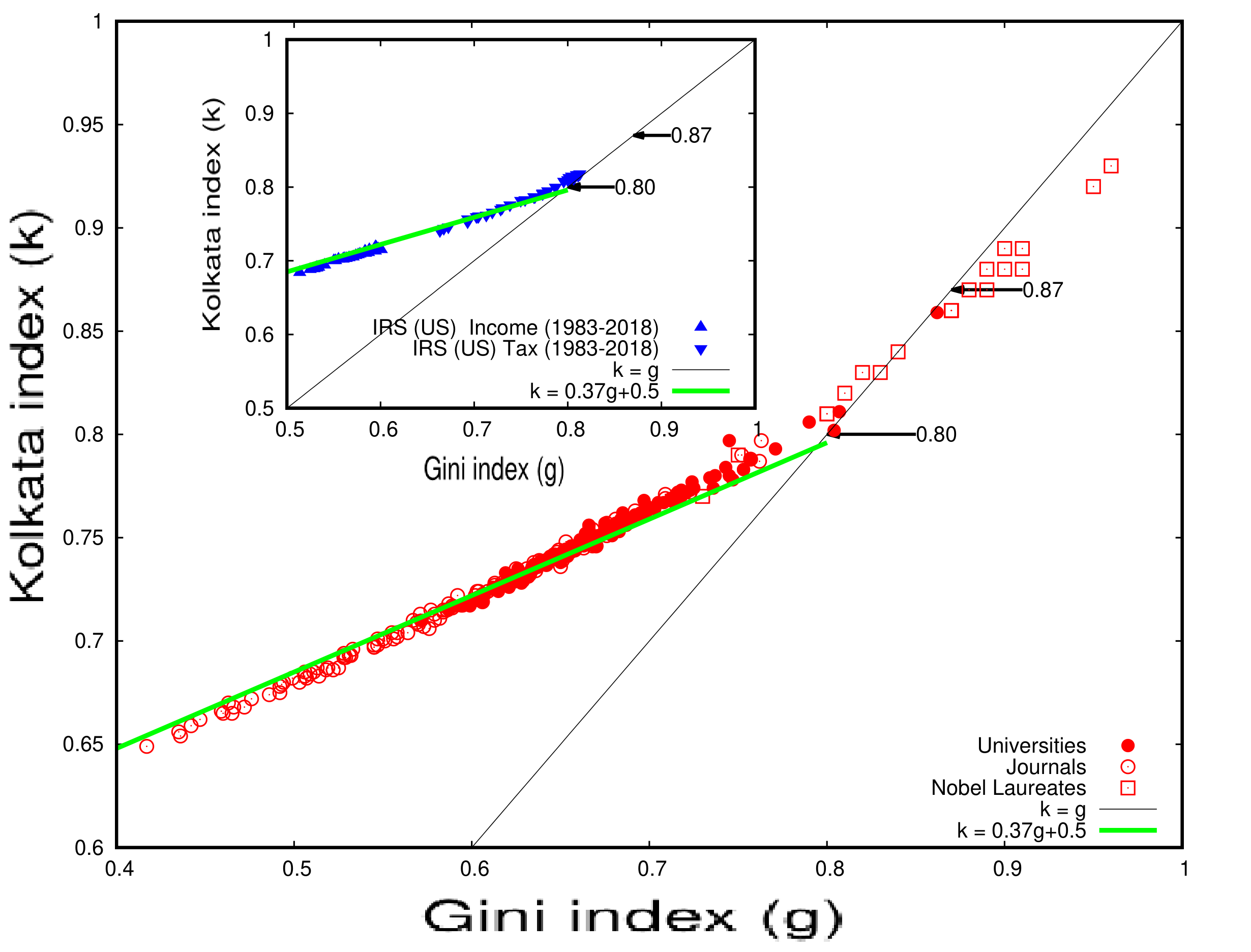}
\caption{Comparison of the Gini index ($g$) and Kolkata index ($k$) obtained from the analysis of the IRS (US) data on income, income tax and income from movies from 1983 to 2018 (see inset in the figure) and citations of papers published by scientists from universities or institutes, published in journals, and Nobel Laureates in Physics, Chemistry, Medicine, and Economics (data taken from Refs. \citep[][]{CG2017} and \citep[][]{Ghosh2021}). The initial variation of $k$ against $g$ for both income and income tax and for citations by universities, journals and individual scientists is remarkably similar and agrees even quantitatively. The main figure illustrates this comparison.(adopted from \citep[][]{Suchi}).} 
\label{fig:9}
\end{figure}

Here we compare the growth pattern of income inequality in the IRS (USA) with citation inequality for papers published by established universities (see tables \ref{tab:4},\ref{tab:5}), published in established journals (see table \ref{tab:6}) and individual Nobel laureate scientists (see table \ref{tab:7}). The data for these comparisons were taken from other publications (Refs. \citep[][]{Chatterjee2016}, \citep[][]{Ghosh2021}, \citep[][]{CG2017}). Figure \ref{fig:9} displays the results, which show that the $k$ and $g$ inequality indices of all these categories drift linearly toward a universal value of $k = g \simeq 0.87$ under unrestricted competition. This suggests that approximately 87\% of the wealth, citations, or votes are possessed, earned, or won by 13\% of people, papers, or election candidates respectively.

\subsection{\textbf{A Study of Inequality in Citations: An Analysis of Individual Authors and Award Recipients}}\label{subsec:5}

In this section, we present two tables with statistical analysis of research papers and their citations for 20 distinguished scientists who have won Nobel Prizes in economics, physics, chemistry, and biology/physiology/medicine, as well as for several international prize winners in mathematics and physics. Table \ref{tab:7} shows the analysis for Nobel Laureates, while Table \ref{tab:8} presents the results for the international prize winners. The data were collected from Google Scholar during the first week of January 2021 and the names of the scientists are reported in the same format as they appear on their respective Google Scholar pages.

\begin{figure}[H]
\centering
\includegraphics[width=15cm]{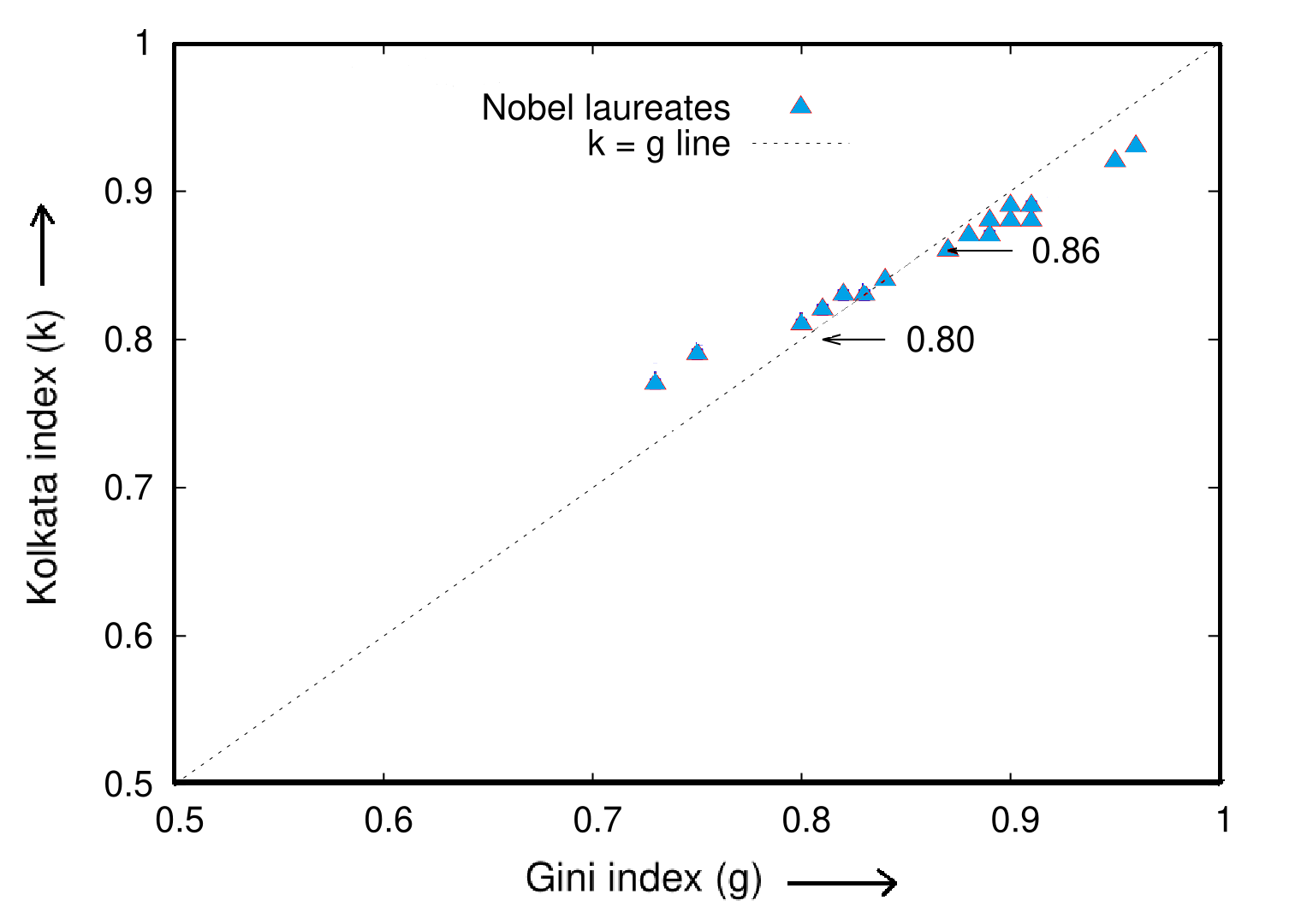}
 \caption{Plot of the inequality index values of Kolkata ($k$) versus the corresponding Gini index ($g$) for the citation statistics of publications by 20 chosen Nobel Laureates, as given in Table \ref{tab:7}. The plot suggests a coincidence value of $k = g = 0.86 \pm 0.06$, as adapted from a previous study \citep[][]{CG2017}.}
 \label{fig:10}
\end{figure}

\begin{figure}[H]
\centering
\includegraphics[width=0.7\textwidth]{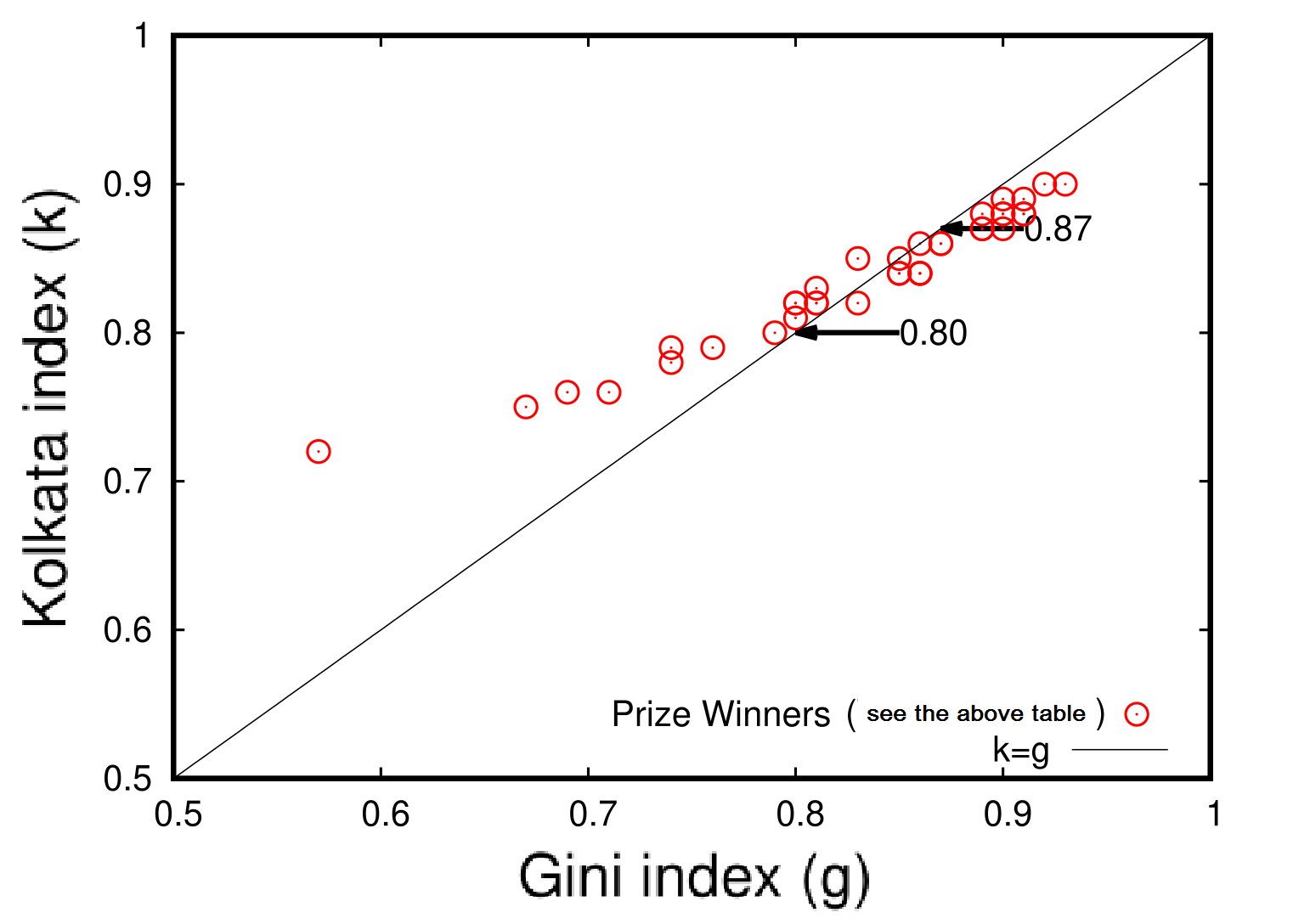}
\caption{Plot of the Kolkata index ($k$) versus the Gini index ($g$) for the citation inequalities in papers published by individual prize winners. The data were extracted from the corresponding Lorenz function $L(p)$ for each scientist and are presented in Table \ref{tab:8}(adopted from \citep[][]{Suchi}).}
\label{fig:11}
\end{figure}

We also provide two figures, \ref{fig:10} and \ref{fig:11}, that illustrate the inequality analysis for the 20 Nobel Laureates and prize winners. These figures demonstrate an ideal example of the dynamics of wealth inequality without any external interventions or fine tuning. Furthermore, the Gini index ($g$) and the Kolkata index ($k$) approach each other at a value of approximately 0.87 in both cases. This indicates that the results are consistent and robust across different groups of distinguished scientists.

\begin{table}[H]
 \centering
\begin{tabular}{|c|c|c|c|c|c|}
\hline
Award & Name of recipients&\multicolumn{4}{c|}{Google Scholar citation data}\\
 \cline{3-6}
 &  & $N_p$ & $N_c$  & \multicolumn{2}{c|}{index values}\\
 \cline{5-6}
 & &  & & $g$ &$k$\\
 \hline
 \multirow{6}{*}{\shortstack[lb]{ NOBEL \\ Prize \\ (Econ.) }}  & Joseph E. Stiglitz & 3000 & 323473 & 0.90 & 0.88 \\
\cline{2-6}
& William Nordhaus & 783 & 74369 & 0.87 & 0.86 \\
\cline{2-6}
& Abhijit Banerjee & 578 & 59704 & 0.89 &0.88 \\
\cline{2-6}
& Esther Duflo & 565 & 69843 & 0.91 & 0.89 \\
\cline{2-6}
& Paul Milgrom & 365 & 102043 & 0.90 & 0.89 \\
\cline{2-6}
& Paul Romer & 255 & 95402 & 0.96 & 0.93 \\
 \hline 
 \multirow{8}{*}{\shortstack[lb]{ NOBEL \\ Prize \\ (Phys.)}} & 
 Hiroshi AMANO & 1300 & 44329 & 0.80 & 0.81 \\
\cline{2-6}
& David Wineland & 720 & 63922 & 0.88 & 0.87 \\
\cline{2-6}
& Gérard Mourou & 700 & 49759 & 0.82 & 0.83 \\
\cline{2-6}
& Serge Haroche & 533 & 40034 & 0.87 & 0.86 \\
\cline{2-6}
& A. B. McDonald & 492 & 20346 & 0.91 & 0.88 \\
\cline{2-6}
& David-Thouless & 273 & 47452 & 0.89 & 0.87 \\
\cline{2-6}
& F.D.M. Haldane & 244 & 41591 & 0.87 & 0.86 \\
 \cline{2-6}
& Donna Strickland & 111 & 10370 & 0.95 & 0.92 \\
\hline 
 
 \end{tabular}
 \begin{tabular}{|c|c|c|c|c|c|}
 \hline
 Award & Name of
recipients  & \multicolumn{4}{c|}{Google Scholar citation data}\\
 \cline{3-6}
  &  & $N_p$ & $N_c$  & \multicolumn{2}{c|}{index values}\\
 \cline{5-6}
 & &  & & $g$ & $k$\\
 \hline 
 \multirow{4}{*}{\shortstack[lb]{ NOBEL \\  Prize \\ (Chem.)}}
 & & & & & \\
 & Joachim Frank & 853 & 48077 & 0.80 & 0.81 \\
\cline{2-6}
& & & & & \\
& Frances Arnold & 682 & 56101 & 0.75 & 0.79 \\
\cline{2-6}
& & & & & \\
& Jean Pierre Sauvage & 713 & 57439 & 0.73 & 0.77 \\
\cline{2-6}
& & & & & \\
& Richard henderson & 245 & 27558 & 0.84 & 0.84 \\
\hline
\multirow{3}{*}{\shortstack[lb]{NOBEL\\ Prize \\ (Bio.)}}
& & & & & \\
& Gregg L. Semenza & 712 & 156236 & 0.81 & 0.82 \\
& & & & & \\
\cline{2-6}
& & & & & \\
& Michael Houghton & 493 & 49368 & 0.83 & 0.83 \\
& & & & & \\
\hline 
 \end{tabular}
\caption{Display the statistical analysis of research papers and their citations for 20 Nobel Laureates in economics (Econ), physics (Phys), chemistry (Chem), and biology/physiology/medicine (Bio). The data were collected from their individual Google Scholar pages with a verifiable email site during the first week of January 2021. To be included in the analysis, the Laureates had to have at least 100 entries (papers or documents), with the latest not before 2018. These Nobel Laureates have a range of papers, from 111 to 3000, with $N_p$ denoting the number of papers. The Laureates' names appear in the same form as they do on their respective Google Scholar pages (Adopted from Ghosh et al. \citep[][]{Ghosh2021}).}
\label{tab:7} 
\end{table}

\begin{table}[H]
 \centering
 \small
\resizebox{\columnwidth}{!}{%
\begin{tabular}{|c|c|c|c|c|c|}
\hline
Award & Name of recipients&\multicolumn{4}{c|}{Google Scholar citation data}\\
 \cline{3-6}
 &  & $N_p$ & $N_c$  & \multicolumn{2}{c|}{index values}\\
 \cline{5-6}
 & &  & & $g$ &$k$\\
% & & $(N_p)$&($N_c)$& &  & \\
 \hline
 \multirow{10}{*}{\shortstack[lb]{ FIELDS \\ Medal \\ (Math.) }}  &Terence Tao & 604 & 80354 & 0.88 & 0.86\\
\cline{2-6}
&Edward Witten&402&314377&0.74&0.79\\
\cline{2-6}
&Alessio Figalli&228&5338&0.67&0.75\\
\cline{2-6}
&Vladimir Voevodsky&189&8554&0.83&0.85\\
\cline{2-6}
&Martin Hairer&181&7585&0.74&0.78\\
\cline{2-6}
&Andrei Okounkov&134&10686&0.69&0.76\\
\cline{2-6}
&Stanislav Smirnov&79&4144&0.76&0.79\\
\cline{2-6}
&Richard E. Borcherds&61&5096&0.81&0.83\\
\cline{2-6}
%&Ngô Bảo Châu&44&1214&17&0.71&0.76\\
&Ngo Bao Chau&44&1214&0.71&0.76\\
\cline{2-6}
&Maryam Mirzakhani&25&1769&0.57&0.74\\
 \hline 
 \multirow{9}{*}{\shortstack[lb]{ASICTP \\ DIRAC \\  Medal \\ (Phys.)}} & Rashid Sunyaev & 1789&103493&0.91&0.88 \\
\cline{2-6}
& Peter Zoller&838&100956&0.81&0.82\\
\cline{2-6}
& Mikhail Shifman&784&52572&0.85&0.84\\
\cline{2-6}
& Subir Sachdev&725&58692&0.83&0.82\\
\cline{2-6}
& Xiao Gang Wen&432&46294&0.8&0.82\\
\cline{2-6}
& Alexei Starobinsky&328&47359&0.81&0.82\\
\cline{2-6}
& Pierre Ramond&318&23610&0.89&0.87\\
\cline{2-6}
& Charles H. Bennett&236&89798&0.9&0.88\\
\cline{2-6}
& V. Mukhanov&208&27777&0.85&0.84\\
\cline{2-6}
& M A Virasoro&150&12886&0.9&0.87\\
\hline
 \end{tabular}
 \begin{tabular}{|c|c|c|c|c|c|}
 \hline
 Award & Name of
recipients  & \multicolumn{4}{c|}{Google Scholar citation data}\\
 \cline{3-6}
  &  & $N_p$ & $N_c$  & \multicolumn{2}{c|}{index values}\\
 \cline{5-6}
 & &  & & $g$ & $k$\\
 \hline 
 \multirow{5}{*}{\shortstack[lb]{BOLTZMANN \\  Award \\ (Stat. Phys.)}}&Elliott Lieb&755&76188&0.86&0.85\\
\cline{2-6}
&Daan Frenkel&736&66522&0.8&0.81\\
\cline{2-6}
&Harry Swinney&577&46523&0.86&0.84\\
\cline{2-6}
&Herbert Spohn&446&25188&0.79&0.8\\
\cline{2-6}
&Giovanni Gallavotti&446&15583&0.86&0.84\\
\hline
\multirow{10}{*}{\shortstack[lb]{JHON Von \\ NEUMANN \\ Award \\ (Social Sc.)}} &Daron Acemoglu&1175&172495&0.91&0.89\\
\cline{2-6}
&Olivier Blanchard&1150&113607&0.91&0.89\\
\cline{2-6}
&Dani Rodrik&1118&136897&0.9&0.89\\
\cline{2-6}
&Jon Elster&885&79869&0.89&0.87\\
\cline{2-6}
&Jean Tirole&717&201410&0.91&0.88\\
\cline{2-6}
&Timothy Besley&632&57178&0.89&0.88\\
\cline{2-6}
&Maurice Obstfeld&586&73483&0.9&0.88\\
\cline{2-6}
&Alvin E. Roth&566&54104&0.87&0.86\\
\cline{2-6}
&Avinash Dixit&557&82536&0.93&0.9\\
\cline{2-6}
&Philippe Aghion&490&119430&0.85&0.85\\
\cline{2-6}
&Matthew O. Jackson&397&39070&0.86&0.84\\
\cline{2-6}
&Emmanuel Saez&310&48136&0.86&0.86\\
\cline{2-6}
&Mariana Mazzucato&236&12123&0.87&0.86\\
\cline{2-6}
&Glenn Loury&226&13352&0.92&0.9\\
\cline{2-6}
&Susan Athey&203&18866&0.8&0.82\\
\hline 
 \end{tabular}
 }
\caption{Citation data analysis for selected prize winners in Physics, Mathematics and Social Sciences. The selected prize winners are recipients of the Dirac Medal, Boltzmann Medal, Fields Medal and John von Neumann Award. The analysis includes only those prize winners who have verified email addresses and an updated Google Scholar page after 2018. The data was collected in the first week of April 2021. For each individual scientist, the table reports the total number of papers ($N_p$), total citations ($N_c$), Gini index ($g$), and Kolkata index ($k$).}
%Citation data analysis for some of the selected Prize winners in Physics (Dirac Medal \& Boltzmann medal) , Mathematics (Fields Medal) and Social Sciences (John von Neumann Award), who have their respective (`Verified email' and updated after 2018) Google Scholar pages (data taken  in the 1st week of April 2021). Here, for each individual scientist, $N_p$ denotes total number of papers, $N_c$ denotes total citations, $g$ denotes Gini index and $k$ denotes Kolkata index.}
\label{tab:8}
\end{table}

In figures \ref{fig:10} and \ref{fig:11} we see that on an average $15\%$ of the papers published by successful individual authors received about $85\%$ of his/her total citations. Also, Gini and $k$-index coincide almost $\simeq 0.87$.

\subsection{\textbf{Similarity in the Behavior of Gini and Kolkata Indices Across Multiple Domains: A Universality Study}}\label{subsec:6}

\begin{figure}[H]
\centering
\includegraphics[width=0.7\textwidth]{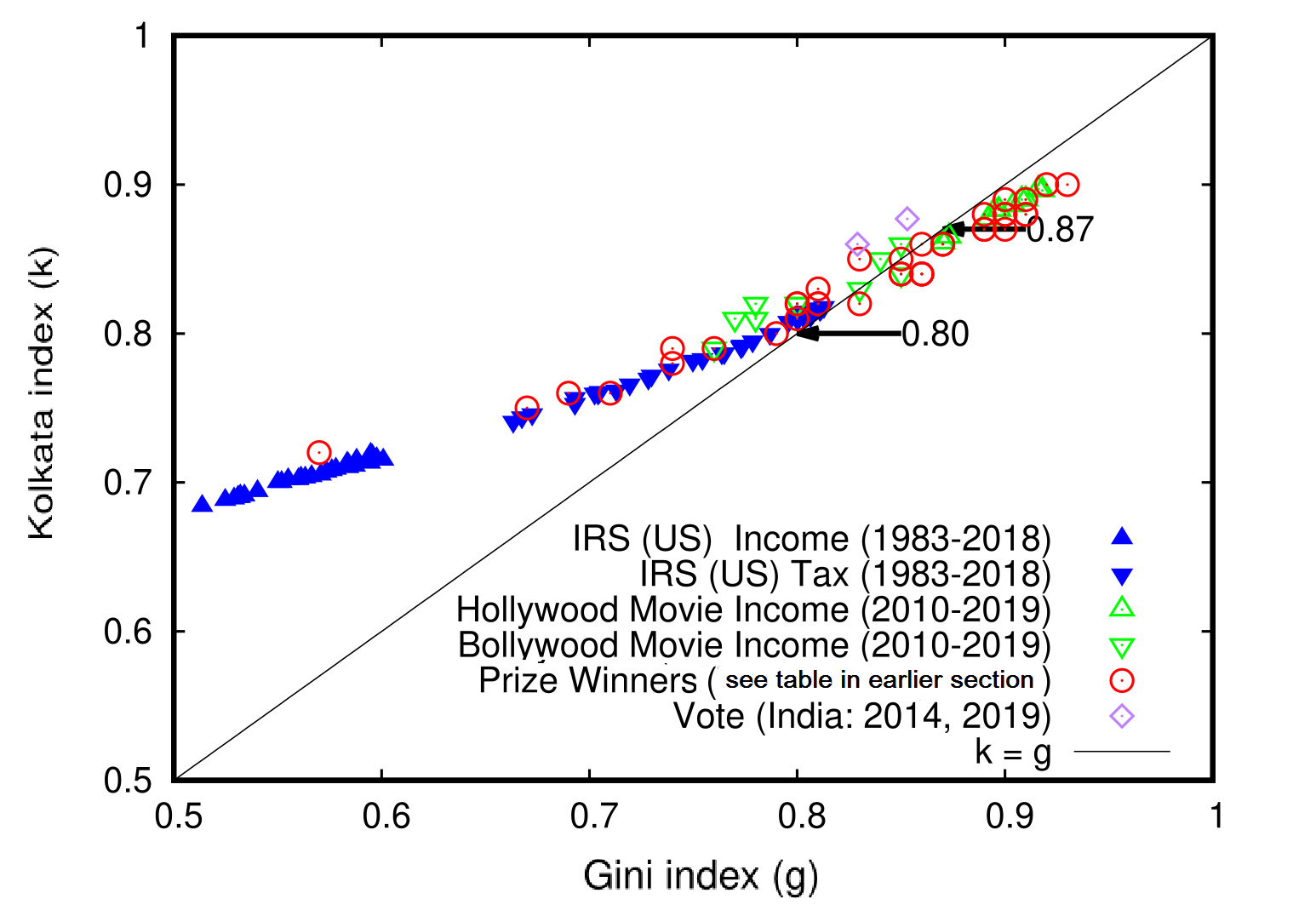}
\caption{A compiled plot of Kolkata index ($k$) values versus corresponding Gini index ($g$) values for several cases analyzed in previous subsections, including household income and income tax data (Figs. \ref{fig:5a}, \ref{fig:X}), movie income (Fig. \ref{fig:5b}), citation inequalities of individual prize winners (Table \ref{tab:8}, Fig. \ref{fig:11}), and vote share inequalities among election contestants (Table \ref{tab:3}). The results suggest that there may be universal inequality measures across social institutions, as the data points in the plot converge towards a common value of $k=g=0.87\pm 0.02$. This observation has important implications for understanding the nature and extent of wealth inequality across different domains of society (adopted from \citep[][]{Suchi}).}
\label{fig:compile}
\end{figure}

In this section, we aim to consolidate the findings on the Gini ($g$) and Kolkata ($k$) indices from previous subsections. We gathered estimates of $g$ and $k$ from various sources, including the IRS (US) data \citep[][]{IRS}, \citep[][]{Ludwig2021} on household income and income tax spanning 1983-2018, the citation data from papers published by 40 international prize winners (Fields medalists, ASICTP Dirac medalists, Boltzmann medalists, and von Neumann awardees), as shown in Table \ref{tab:8}, and the vote share data from the Indian Parliament elections in 2014 and 2019, displayed in Table \ref{tab:3}. We present the compiled results in Fig. \ref{fig:compile}. The collective analysis of all these results reveals a universal trend of inequality growth across various social institutions, markets (income and wealth), academic institutions (citations), and elections (vote shares among the candidates). Furthermore, the analysis shows that the measures of inequality converge to $k=g=0.87\pm0.02$.

\subsection{\textbf{Inequality analysis for Man-made conflicts and Natural Disasters}}\label{subsec:7}
We will see that in man-made conflicts like war, battle, armed-conflict, terrorism, murder etc., on an average for $85\%$ of human deaths are caused by $15\%$ of social conflicts including war (see Table \ref{tab:man-made}).

 \begin{table}[H]
 \centering
     \begin{tabular}{|l|c|c|} 
    \hline
      \textbf{Type of conflicts} & \textbf{$g$-index} & \textbf{$k$-index} \\
      \hline
		war & $0.83\pm$0.02 & $0.85\pm$0.02  \\
		\hline     
		battle & $0.82\pm$0.02 & $0.85\pm$0.02 \\
		\hline
      	  armed-conflict & $0.85\pm$0.02 & $0.87\pm$0.02  \\
      \hline
          terrorism & $0.80\pm$0.03 & $0.83\pm$0.02  \\
      \hline
       murder & $0.66\pm$0.02 & $0.75\pm$0.02 \\
		\hline	   
    \end{tabular}
 \caption{Estimated inequality index values for Gini ($g$) and Kolkata ($k$) applied to man-made conflicts, such as wars and acts of terrorism. The data shown is adapted from a previous study (\citep[][]{Sinha2019}) and represents death counts as a measure of inequality.}
\label{tab:man-made}
\end{table}
 
We will also show that for natural disasters like earthquake, flood, tsunami etc, almost $95\%$ of human deaths are caused by $5\%$ of the disasters (see Table \ref{tab:natural}).

\begin{table}[H]
\centering
    \begin{tabular}{|l|c|c|} 
    \hline
      \textbf{Type of disasters} & \textbf{$g$-index} & \textbf{$k$-index}\\
      \hline
		  earthquake & $0.94\pm$0.02  & $0.95\pm$0.02\\
		\hline     
		 flood & $0.98\pm$0.02 & $0.98\pm$0.02\\
		\hline
      	   tsunami & $0.93\pm$0.02 & $0.94\pm$0.02\\
    \hline	   
    \end{tabular}
    \caption{Estimated values of the Gini index ($g$) and Kolkata index ($k$) as measures of inequality in death counts resulting from natural disasters such as earthquakes and tsunamis (adapted from \cite{Sinha2019}). The data presented here provide valuable insights into the distribution of fatalities resulting from natural disasters and the effectiveness of these inequality indices in capturing the severity of such events.}
    \label{tab:natural}
\end{table}

\subsection{\textbf{Inequality analysis in Computing systems}}\label{subsec:8}
The field of computer science has long recognized the adage that "$20\%$ of the code contains $80\%$ of the errors", as noted in \citep[][]{Wiki_comp}. Consequently, software developers have a vested interest in identifying and rectifying this critical $20\%$ of the codebase in order to enhance the quality of the software. In a related finding, researchers have also observed that approximately $80\%$ of the functionality of a given software program can typically be implemented in just $20\%$ of the total development time. Conversely, the remaining $20\%$ of the software's features, which represent the most challenging and time-consuming aspects of the coding process, often require the remaining $80\%$ of the total development time. This factor is commonly taken into account when estimating the cost and timeline for software development using the Constructive Cost Model (COCOMO) approach.
So in computing we see that $20\%$ of the code has $80\%$ of the errors.

\subsection{\textbf{ Inequality analysis for sports: Olympic medals share}}\label{subsec:9}
In recent discussions, scholars have posited that the concept of inequality also applies to the field of sports, where a few top performers often dominate the majority of victories. This phenomenon is exemplified in the sport of baseball, where Wins Above Replacement (WAR) is used as a composite metric to gauge a player's overall importance to a team. Recent statistical analyses have revealed that a mere $15\%$ of baseball players accounted for $85\%$ of the total wins, while the remaining $85\%$ of players were responsible for generating only $15\%$ of the wins \citep[][]{Wiki_Sports_1}. These findings provide compelling evidence for the existence of significant inequality within the sport, indicating that a small minority of players are driving the majority of team successes.
The inequality statistics (reflected in the Gini and Kolkta index values) for the country-wise inequalities in winning Olympic medals are shown in Table \ref{oly}. Typically 16 to 13 percent of countries win 84 to 87 percent  of the Olympic medals (gold, silver, or bronze) in the  Summer Olympics (see Table \ref{oly}; statistics for the last four Olympics from ref. \citep[][]{Wiki_Sports_2}).

\begin{table}[H]
\centering
\begin{tabular}{|c|c|c|c|}
\hline
Year & Medal & $g$ & $k$ \\
\hline
\multirow{4}{*}{2020}  &Gold& 0.87 & 0.85 \\
\cline{2-4}
&Silver& 0.86 & 0.85\\
\cline{2-4}
& Bronze & 0.84 & 0.84\\
\cline{2-4}
&Total & 0.84 & 0.83 \\
\hline
\multirow{4}{*}{2016}  &Gold& 0.88 & 0.87\\
\cline{2-4}
&Silver& 0.86 & 0.85 \\
\cline{2-4}
& Bronze & 0.85 & 0.85 \\
\cline{2-4}
&Total & 0.85 & 0.84 \\
\hline
\end{tabular}
\begin{tabular}{|c|c|c|c|}
\hline
Year & Medal & $g$ & $k$ \\
\hline
\multirow{4}{*}{2012}  &Gold& 0.89 & 0.87 \\
\cline{2-4}
&Silver& 0.87 & 0.85\\
\cline{2-4}
& Bronze & 0.84 & 0.84 \\
\cline{2-4}
&Total & 0.85 & 0.85 \\
\hline
\multirow{4}{*}{2008}  &Gold& 0.89 & 0.87 \\
\cline{2-4}
&Silver& 0.85 & 0.84 \\
\cline{2-4}
& Bronze & 0.86 & 0.85 \\
\cline{2-4}
&Total & 0.85 & 0.84 \\
\hline
\end{tabular}
\caption{Inequality statistics  of the Olympic medals and the winning countries (2008 - 2020). Typically the medal number (for gold, silver or bronze, each) is 300 and contesting countries are about 200. When the cumulative fraction of medals is plotted  against the fraction of countries winning them (ordered from the least medal winning to highest), one gets the Lorenz curve and we estimate the Gini ($g$) and Kolkata ($k$) indices for each  year. As competitions are extremely high in the Olympics (with no welfare support for equality in achievements among contestants), typically we find 16 to 13 percent of countries win 84-87 percent  of medals (gold, silver or bronze) in any Olympic.}
\label{oly}
\end{table}

\section{ Growing avalanche size inequalities in
Sand-pile models: Universality near the SOC point}\label{sec:6}

With all the observations made till now we find that the value of $k$-index reaches a critical value of $k=g$ before the inequality falls again.
Such kind of observation implies that $k=g$ point can be characterized as a critical point of the society.
Further observations with physical self-organized critical systems has also indicated same kind of behavioural pattern \citep[][]{Manna}.
As all of the above observations were made for unrestricted competitive scenarios where only the winners get all the facilities, we can expect that in real society such kind of point exists and the inequality can never reach beyond that point because of the government subsidies which aid the poor population.

According to the self-organized criticality (SOC) framework, the critical point is an attractor, and we found (see for example \citep[][]{Manna}) just preceding the SOC point, the avalanche size inequality attain a value approximately equal to $0.86$. This observation is consistent with our observations discussed in the previous sections.

\begin{figure}[H]
\begin {center}
\includegraphics[scale=0.4]{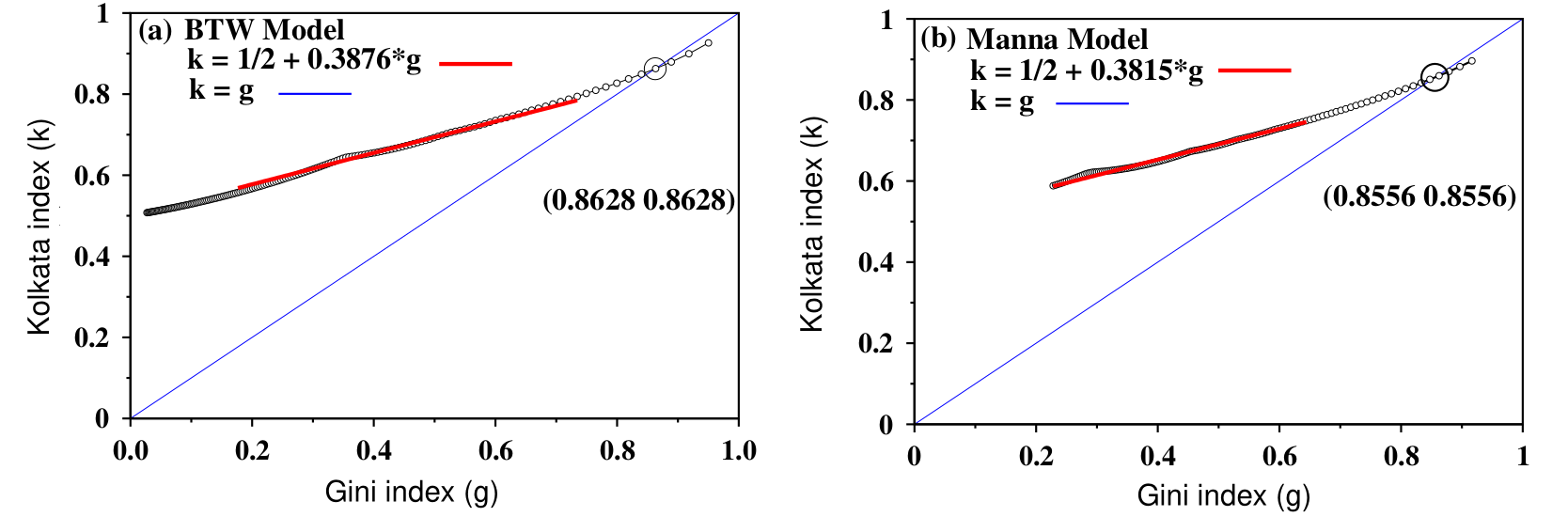}
\end {center}
\caption{The $k$-index versus Gini index ($g$) relationship for two sandpile models: (a) the BTW model, and (b) the Manna model. In both cases, the initial portions of the curves follow a straight line with slightly different slopes, as demonstrated in the figures. The crossing points of the curves with the line $g=k$ are found to be 0.8628 and 0.8556 for the BTW and Manna models, respectively. This figure is adapted from the work of \cite{Manna}.}
\label {g_k_sandpile}
\end{figure}

We present here the results of a comparison of the $k$ and $g$ indices for two different sandpile models, namely the Bak-Tang-Wiesenseld (BTW) model and Manna model \citep[][]{Manna}. Fig.~\ref{g_k_sandpile} illustrates the $k$ versus $g$ relationship for these models, with panel (a) showing the results for the BTW model and panel (b) for the Manna model. The plots depict a linear relationship between $k$ and $g$ for the initial part of the curve. For the BTW model, the slope of the linear fit is found to be 0.3876, while for the Manna model, the slope is slightly lower at 0.3815. The intersection of the plots with the line of equality ($k=g$) occurs at 0.8628 and 0.8556 for the BTW and Manna models, respectively. It is important to note that these results were obtained for a system size of $L \times L = 512 \times 512$.
 
\begin{figure}[H]
\begin {center}
\includegraphics[scale=0.7]{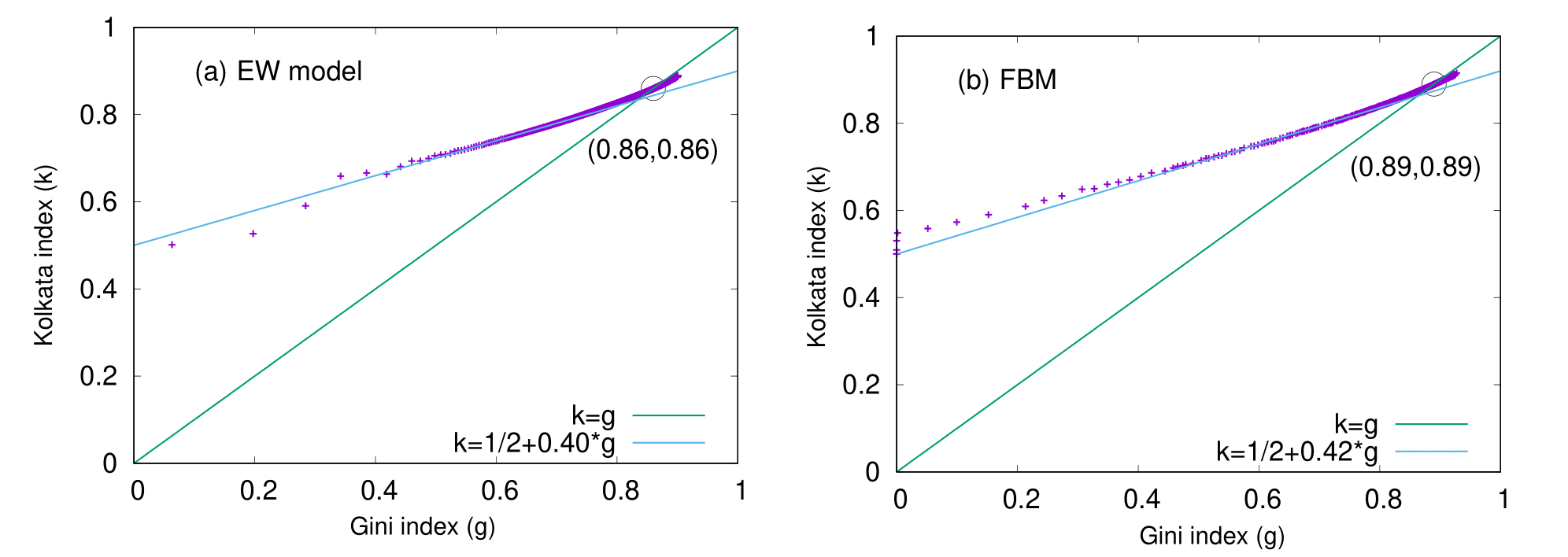}
\end {center}
\caption{The relationships between the $k$-index and Gini index ($g$) for the EW and centrally loaded fiber bundle models. Fig. (a) shows the $k$ versus $g$ plot for the EW model, with an initial slope of 0.40. Fig. (b) displays the same plot for the centrally loaded fiber bundle model, with an initial slope of 0.42. The figure is adapted from \citep[][]{Manna}.}
\label{k_vs_g_EW_fbm}
\end{figure}

The relationship between the Kolkata index ($k$) and Gini index ($g$) is also analysed here for two other SOC models, mainly Edwards-Wilkinson (EW) model and the Fibre-Bundle Model (FBM) \citep[][]{Manna} in this section. In Fig.\ref{k_vs_g_EW_fbm}(a), the $k$ values are plotted against their corresponding $g$ values for the EW model. A linear relationship is observed in the initial section of the plot, with a slope of 0.40. Similarly, in Fig.\ref{k_vs_g_EW_fbm}(b), the $k$ versus $g$ plot is shown for the centrally loaded fiber bundle model. In this case, the initial linear slope is measured to be 0.42. These results suggest that the $k$-index and Gini index are linearly related in the early stages of both models, with slightly different slopes.

In several SOC models considered here, we observe a remarkably consistent coincidence of the avalanche size inequality indices ($g$ and $k$) at around $g = k \simeq 0.86$ just preceding the arrival of the SOC point. This is also very consistent with what we have observed earlier in the inequality measures in various socioeconomic contexts. We consider this to be noteworthy.

\section{Summary and Discussions}

Disparities between social classes are constantly present (see, e.g., \cite{Jose,Boghosian} for some recent discussions), which has been proposed that it is an emergent trait of complex socioeconomic systems \cite{zhukov} with many interacting parts.
It may be mentioned at this point that allowing for a higher probability of exchange for the poor in conservative kinetic exchange models induces (see, e.g.,~\cite{Pian,Iglesias,Pian1}), in a novel, self-organized way, a minimum ``poverty level'', thereby reducing inequality. Herein, we attempted to show that the extreme social inequality we find  in society is the result of built-in self-organized critical dynamics. 
A self-organized criticality \cite{Watkins,cite_soc} framework has been proposed to explain the evolutionary behavior of diverse systems, such as wealth distributions, financial markets, cryptocurrencies, citation dynamics, etc. However, a~major unresolved issue concerns the appropriate quantification of the observed disparities and their potential universality across systems. Pareto's 80/20 law, which implies that 80\% of the wealth ends up in the hands of 20\% of the richest ($k=0.80$) members of society, has traditionally been used as a benchmark for measuring the degree of extreme social inequality. Further, we attempted to determine the amount of social and economic inequality in a number of very competitive systems without any outside interventions to stop halt inequality among the agents.
This self-tuning feature of competitive dynamics in various social sectors suggests similar inequality behavior in the SOC system. Moreover, we studied a significant number of socioeconomic systems through the framework of SOC architecture. This approach has been used to analyze a range of systems, including financial markets \cite{markets_soc}, citation evolution \cite{cite_soc,TS}, cryptocurrencies \cite{bitcoin_soc}, and~political behavior \cite{politics_soc}, among~others. Despite the diversity of these systems, our analysis reveals that the behavior of inequality indices, particularly the Gini ($g$) and Kolkata ($k$) indices, demonstrates near-universal characteristics across socioeconomic systems. Specifically, we observe that these indices tend to converge towards a value of approximately 0.87. This finding is particularly noteworthy, given that similar behavior has been observed in SOC models of physical systems \cite{Manna,front}. A recent research publication \cite{Thurner} presented findings that highlight the high level of inequality in group size distribution. The~authors reported a $g$ index value of $0.90$ for both the theoretical and empirical observations of group size distribution. We note the consistency of such a high value of the $g$ index with our~observations.

In Section~\ref{sec:2}, we presented a chronological account of the development of social inequality measures since 1896 and demonstrated their similarity to those in sand pile models immediately prior to their respective self-organized critical (SOC) points. In Section~\ref{sec:3}, we discussed the Lorenz curve, the Gini ($g$) index, and the Kolkata ($k$) index in detail and presented proofs of their properties, as well as exemplary calculations.
In Section~\ref{sec:4} of our review, we conducted an investigation into several analytical characteristics of the Lorenz function ($L(p)$). Our analysis, supported by Tables~\ref{tab:1} and \ref{tab:2}, led us to the conclusion that for~a wide range of plausible analytic forms of $L(p)$, the~values of the Gini ($g$) and Kolkata ($k$) indices that correspond to coincidence points fall within the range of $4/5$ (equivalent to $0.80$) to $8/9$ (equivalent to $0.88\dots$). This finding suggests that the values of $g$ and $k$ are tightly constrained and~are influenced by the analytic properties of the Lorenz function ($L(p)$). In particular, in~Section~\ref{sec:5}, we considered datasets accounting for factors such as earnings from different sectors, Bitcoin price fluctuations, citations of selected prize-winning scientists, and votes received by various candidates in elections. We conducted an analysis of income data from multiple sources. Specifically, we examined data from the Internal Revenue Service (IRS) in the United States, as~well as income tax data, over~a period of 36 years spanning 1983 to 2018. The~IRS data were obtained from previous research \cite{IRS, Ludwig2021}. Additionally, we investigated income data from the Hollywood movie industry in the United States, which was sourced from a prior study \cite{Hollywood2011}, and~data from the Bollywood movie industry in India \cite{Bollywood2011} for the time period ranging from 2011 to 2019. These data sources were analyzed in detail in Section \ref{subsec:1} of our study. In Section \ref{subsec:2}, we showed data for Bitcoin price fluctuations \cite{Bitcoin2021}, and in Section \ref{subsec:3}, we examined the vote share data pertaining to candidates who ran in parliamentary elections in India during in 2014 and 2019. We conducted an analysis of this data with the aim of exploring inequalities in the vote share among candidates \cite{Lokesabha2014,Lokesabha2019}.
In this research, we present data on the citations of published papers from several prominent universities and institutions, as~well as leading journals, which can be found in Section \ref{subsec:4} and are cited in~\cite{k-index}. We also conducted an analysis of citation data from Google Scholar for 20 selected individuals who have been awarded Nobel Prizes in the fields of economics, physics, chemistry, and~biology/physiology/medicine, as~well as individuals who have been awarded the Fields Medal (mathematics), the Boltzmann Medal (physics),the ASICTP Dirac Medal (physics), or the~John von Neumann Award (social science) in various years. We focused on individuals who have their own Google Scholar pages with ``verified email'' addresses; the results of our analysis are presented in Section~\ref{subsec:5}.
We utilized these data sources to investigate sectoral inequality and computed the associated Gini ($g$) and Kolkata ($k$) indices. The~results of our analysis are presented in Figures~\ref{fig:5a}, \ref{fig:5b} and \ref{fig:6b}--\ref{fig:11} and~summarized in Tables~\ref{tab:h_b}--\ref{tab:8}. We also compiled these findings into a single figure, shown in Section~\ref{subsec:6} as Figure~\ref{fig:compile}. Our analysis revealed a universal value of approximately 0.87 for the coinciding $g$ and $k$ indices, indicating an emerging trend of increasing disparities under conditions of competition.
Moreover, in~Section \ref{subsec:7}, we investigated a similar trend for manmade conflicts such as war, terrorism, etc., as well as for natural disasters such as earthquakes, tsunamis, etc. (see Tables~\ref{tab:man-made} and \ref{tab:natural}).
In Sections \ref{subsec:8} and \ref{subsec:9}, we discussed the universality of the $k$ index for computing systems and sports, respectively. The results of our empirical investigation reveal a consistent pattern in the dynamical behavior of the Kolkata index ($k$) and the Gini index ($g$) across various scenarios. Specifically, our findings indicate that these inequality measures converge towards a universal value of $k = g = 0.87 \pm 0.02$ in~situations in which competitions are not subject to any restrictions. This trend was observed consistently across all socioeconomic systems, highlighting the robustness and universality of this phenomenon.
When we talk about dynamics, we are referring to the long-term changes and eventual saturation brought about in the aforementioned systems. We presented a graphical representation of the $g$ and $k$ indices for the daily price fluctuations of Bitcoin over the period of a decade from 2010 to 2021. As~shown in Figure~\ref{fig:6b}, our findings suggest that the $g$ and $k$ indices tend to stabilize at a value of approximately 0.87, which lends further support to our conclusions. This pattern suggests that without the intervention of a central bank, such as in the case of national currencies, the~inequality indices for cryptocurrencies converge. Specifically, both $g$ and $k$ approach a value of 0.87 before decreasing. Our results indicate that the daily swings in the price of Bitcoin, on~average, do not exceed this limiting value ($g = k \simeq 0.87$). Table~\ref{oly} (Section \ref{subsec:9} on Inequality Analysis for sports: Olympic medal share) shows that the $k$ index value typically ranges from 0.84 to 0.87, implying that 13 to 16 percent of countries win 84 to 87 percent of Olympic~medals.
 
We investigated the behavior of inequality indices over time and found that they have not yet converged to the predicted attractor value of 0.87 that results from a self-organized critical (SOC) state. However, our analysis presented in Figure~\ref{fig:X} in Section \ref{subsec:1} reveals that both the Gini index ($g$) and the Kolkata index ($k$) have exhibited a steady increase over time. This trend is likely attributable to the gradual reduction in public welfare programs in the United States. Interestingly, our results demonstrate that the Pareto value of $k=0.80$ has already been surpassed. It is possible to estimate that it will reach 0.87 if all of the aforementioned public assistance programs are eliminated, thereby allowing participants to enter a state of unrestricted competition. In Section~\ref{sec:6}, we discussed the SOC state of different physical systems (such as the BTW model and Manna model) and calculated their inequality indices ($g$ and $k$) in terms of their growing avalanche sizes, showing the universality trend, i.e., $g=k \simeq 0.86$ (Figures~\ref{g_k_sandpile} and \ref{k_vs_g_EW_fbm}).  
 
The social dynamics of competition take the index values of $g = k \simeq 0.87$, indicating that roughly 87\% of wealth, citations, votes, or Olympic medals are possessed, earned, or won by 13\%  of the population, papers, election candidates, or (Olympic participant) countries, respectively in cases of unrestricted competition in which no welfare support towards equality is available.
This may be a quantitative and universal (across all social sectors) version of the 80/20 law ($k=0.80$) observed by Pareto more than a century ago. This property of the inequality indices is intrinsic to the SOC character of the underlying dynamics, and~it has been demonstrated to be present in a wide variety of SOC models in physical science \cite{Manna,front}.

Previous studies \cite{bijin, biro} established that the Gini index ($g$) can be considered to represent the information entropy of social systems, while the Kolkata index ($k$) can be thought of as a representation of the inverse of the effective temperature of such systems. An~increase in $k$ corresponds to a decrease in the average wealth of a society in circulation, resulting in a decrease in temperature. In~this study, we observed that the ratio of $g/k$, which is equivalent to free energy, displays an identical value at multiple points ($g=k\simeq0.87$ and $g=k=1$). These findings suggest the existence of a first-order-like phase transition \cite{Ghosh2021} at the point at which $g=k\simeq0.87$. This reinforces the idea that the relationship between the Gini and Kolkata indices can be analyzed from a thermodynamic perspective. This inequality growth is entropy-driven, as~conjectured in the context of self-organized sand pile systems (see, e.g.,~\cite{main,lang}), similar to those explored herein.

\section{Acknowledgment}

We are grateful to Sai Krishna Challagundla, Arnab Chatterjee, Nachiketa Chattopadhyay, Suhaas Reddy Guntaka, Bijin Joseph, Hanesh Koganti, Anvesh Reddy Kondapalli, Raju Maiti,  Subhrangshu Sekhar Manna, Suresh Mutuswami, Dachepalli R. S. Ram, and Antika Sinha for collaborations at various stages of this study.
BKC is grateful to the Indian National Science Academy for the award of a Senior Scientist Research Grant.
SB is thankful to the DST, Government of India, for financial support of INSPIRE fellowship.
SJB acknowledges using HPCC Surya computing facility in SRM University - AP.

\end{document}